\documentclass[useAMS,usenatbib]{mn2e}
\usepackage{aas}
\usepackage{graphicx}
\usepackage{amsmath}
\usepackage{mathptmx}
\def\idm#1{{\mbox{\scriptsize #1}}}

\def\tv#1{{\pmb #1}}

\newcommand\mjup{m_{\idm{Jup}}}
\newcommand\Chi{\chi^2_\nu}
\usepackage{color}
\definecolor{myred}{rgb}{0.55,0.05,0.05}
\definecolor{myblue}{rgb}{0.2,0.0,0.7}
\definecolor{mybrown}{rgb}{0.9,0.4,0.3}
\newcommand\ednote[1]{{\color{myred}\em\bfseries #1}}
\newcommand\corr[1]{{\color{myred}\sf #1}}
\renewcommand\corr[1]{{\color{black} #1}} 

\newcommand\hide[1]{}
\author[Go\'zdziewski et al.]{%
K.~Go\'zdziewski$^1$, A.~S\l{}owikowska$^2$, D.~Dimitrov$^3$, 
K.~Krzeszowski$^2$, M.~\.Zejmo$^2$, \newauthor 
G.~Kanbach$^4$, V.~Burwitz$^4$, A.~Rau$^4$,
P.~Irawati$^5$, A. Richichi$^5$, M. Gawro\'nski$^1$, \newauthor
G. Nowak$^{6,7,1}$, I.~Nasiroglu$^8$ \& D.~Kubicki$^1$\\\\
$^1$ Toru\'n Centre for Astronomy, Faculty of Physics, Astronomy and Applied
Informatics, N.~Copernicus University, Grudziadzka 5, 87-100 Toru\'n, Poland\\
$^2$ Kepler Institute of Astronomy, University of Zielona G\'ora, Lubuska 2,
65-265 Zielona G\'ora, Poland\\
$^3$ Institute of Astronomy, Bulgarian Academy of Sciences, 72 Tsarigradsko
Chausse Blvd., 1784 Sofia, Bulgaria\\
$^4$ Max Planck Institute for Extraterrestrial Physics, Giessenbachstrasse,
85748 Garching, Germany\\
$^5$ National Astronomical Research Institute of Thailand, 191 Siriphanich
Bldg., Huay Kaew Rd., Suthep, Muang, Chiang Mai 50200, Thailand\\ 
$^6$ Instituto de Astrof\'isica de Canarias, C/ v\'ia L\'actea, s/n, E-38205
La Laguna, Tenerife, Spain\\
$^7$ Departamento de Astrof\'isica, Universidad de La Laguna, Av. Astrof\'isico Francisco S\'anchez, s/n, E-38206 La Laguna, Tenerife, Spain \\
$^8$ Department of Physics, Faculty of Science, Atat\"urk University, Erzurum,
25240, Turkey
}
\begin{document}
\date{\today}
\pagerange{\pageref{firstpage}--21} \pubyear{2013}
%
\title{The HU Aqr planetary system hypothesis revisited}
%
\maketitle
\label{firstpage}
%
\begin{abstract}
We study the mid-egress eclipse timing data gathered for the cataclysmic
binary HU~Aquarii during the years 1993--2014. The (O-C) residuals were
previously attributed to a single $\sim 7$ Jupiter mass companion in $\sim
5$~au orbit or {to a stable 2-planet system with an unconstrained outermost
orbit. We present 22~new observations gathered between June, 2011 and July,
2014 with four instruments around the world. They reveal a systematic
deviation of $\sim 60$--$120$~seconds from the older ephemeris}. {We
re-analyse the whole set of the timing data available. Our results {provide an
erratum to the previous HU Aqr planetary models}, indicating that the
hypothesis for a third and fourth body} in this system is uncertain. The
dynamical stability criterion and a particular geometry of orbits rule out
coplanar 2-planet configurations. {A putative HU Aqr planetary system may be
more complex, e.g., highly non-coplanar. Indeed, we found examples of 3-planet
configurations with the middle planet in a retrograde orbit, which are stable
for at least 1~Gyr, and consistent with the observations. The (O-C) may be
also driven by oscillations of the gravitational quadrupole moment of the
secondary, as predicted by the Lanza et al. modification of the Applegate
mechanism.} Further systematic, long-term monitoring of HU~Aqr is required to
interpret the (O-C) residuals.
\end{abstract}
%
\begin{keywords}
   extrasolar planets---R\o{}mer effect---cataclysmic variables---star: HU~Aqr
\end{keywords}
%
\section{Introduction}
\label{sec:intro}
The HU~Aquarii binary system (HU~Aqr from hereafter) is one of the brightest
polars discovered so far \citep{Schwope1993, Warner1995}. This binary belongs
to the class of magnetic cataclysmic variables (CVs). It consists of a
strongly magnetised white dwarf (WD, primary component) and a main--sequence
red dwarf (RD, M4V spectral type, secondary component). The RD fills its Roche
lobe \citep{Warner1995}. The spin periods of both components are synchronised
with the short orbital period of $\sim 125$~min by tidal forces and extremely
strong magnetic fields. Since the discovery in 1991 \citep{Schwope1993},
eclipses of HU~Aqr have been monitored frequently by several groups and many
different facilities. Because the mid-egress phase of the eclipses is very
short, spanning a few seconds only, it is widely adopted in the literature as
a standard eclipse time-marker. Already a decade ago, \cite{Schwope2001}
reported deviations of the eclipse timing from the linear ephemeris (O-C).

A~few physical phenomena intrinsic to the binary system, like magnetic
braking, mass transfer \citep{Schwarz2009,Vogel2008}, Applegate cycles
\citep{Applegate1992}, orbital precession of the binary \citep{Parsons2014},
the rigid body precession of the WD \citep{Tovmassian2007} and a migration of
the accretion spot on the WD surface are commonly ruled out to be the source
of the (O-C) for HU Aqr \citep{Vogel2008,Schwarz2009,Qian2011}, see also the
most recent analysis by \cite{Bours2014} and references therein. Instead, the
(O-C) variations have been interpreted as the light travel time effect (LTT
from hereafter, or the R\o{}mer effect) due to the presence of one or more
massive, Jovian companions
\citep{Schwarz2009,Qian2011,Hinse2012,Gozdziewski2012}. A common problem of
multiple-planet models is their short-time dynamic instability spanning just
10$^3$---10$^4$ yr time--scales \citep{Horner2011,Hinse2012,Wittenmyer2012}. This
indicates that the common understanding of mechanisms driving the (O-C) in HU
Aqr may be in fact incomplete. Indeed, \cite
{Lanza1998,Lanza1999,Lanza2002,Lanza2004,Lanza2005,Lanza2006} derived
a~mechanism of the orbital period modulations in close binaries due to
magnetic activity cycles in one component, extending the idea of
\cite{Applegate1992}. This theory is build upon a hypothesis that the action
of the hydromagnetic dynamo and the Lorentz force in the convective zone of
the active star may cyclically change its quadrupole moment. This is
sufficient for inducing the orbital period variability with only a~fraction of
the energy required by the simplified Applegate approach
\citep{Lanza1998,Lanza2006}, see also a note in \cite{Brinkworth2006}.

In our previous paper \citep{Gozdziewski2012}, we demonstrated that planetary
models of the (O-C) may be affected by a non-proper,
kinematic formulation of the (O-C). Kinematic (Keplerian) models are
unsuitable for strongly interacting, massive planets, presumably close to
low-order mean motion resonances (MMRs). This is known at least since the
remarkable paper by \citet{Lauglin2001} devoted to the Radial Velocity (RV)
discoveries of extrasolar planets. The LTT and RV models have in fact a very
similar mathematical formulation and concern similar mass ranges and orbital
scales of planetary systems. Following this idea, in \citet{Gozdziewski2012},
we introduced the self-consistent $N$-body model of the LTT effect. This
Newtonian formulation revealed a continuum of {\em stable, 2-planet
configurations} of the HU~Aqr system with an unconstrained outermost orbit. 

We found that the parabolic ephemeris permitting such stable solutions must
involve an excessively large secular decrease of the binary period. Only for
stable 2-planet configurations involved in low order MMRs, like the 3:2~MMR,
spanning narrow islands in the orbital elements space, we found the orbital
period decrease to be reasonably small, though still 2-3 times larger than it
is usually explained by magnetic braking, mass transfer or Applegate cycles,
as argued by \cite{Vogel2008,Schwarz2009,Qian2011,Bours2014}; see however the
note above and discussion in this paper. We suggested a possible solution of
this paradox by selecting a homogeneous subset of light-curves in the optical
domain. We proposed that non-unique 2-planet models in the literature might
appear due to mixing timing data in different spectral domains (infrared,
ultraviolet, $X$--rays). That proposition was reinforced by observations
derived from the fast photometer OPTIMA. The OPTIMA photometry in the optical
domain, spanning more that 10~years between 1999-2012, exhibits formal
sub-second accuracy and the collected data revealed an apparently perfect,
single period (O-C) variation of $\sim 40$~s full amplitude. Combined with
other optical measurements, it might be attributed to a single, massive Jovian
planet of $\sim 7\,\mjup$ in a $\sim 5$~au orbit, simultaneously minimising
the number of free parameters in the (O-C) model. However, shortly towards the
end of the observing season 2012, we noticed significant deviations from the
parabolic ephemeris in \citet{Gozdziewski2012}. After observations during
September, 2013, it was already clear that the (O-C) exceed the LTT
semi-amplitude of all the previous planetary models by more than a factor of
two, also ruling out the 1-planet solution proposed in \cite{Gozdziewski2012}.
Unfortunately, our 1-planet (single-period) solution over-emphasises the
subset of the optical data, and the timing variability due to spectral windows
and different geometry of the eclipses appears as non-significant. 

Therefore, we found it necessary to conduct a new analysis of the up-to date
set of all observations and to revise all LTT models derived so far
\citep{Schwarz2009,Qian2011,Hinse2012,Gozdziewski2012}. Since the $N$-body
2-planet models in \citet{Gozdziewski2012} were constructed on the basis of
{\em all} available data, we continue the previous work, \corr{providing} also
an erratum to this paper. We stress here, that we test the LTT hypothesis in
detail, as one of possible causes of the (O-C) variability, recalling that the
\cite{Lanza2006} theory might also explain the (O-C), without invoking the
R\oe{}mer effect at all.
 
During corrections of this manuscript in accord with the reviewer's report, we
found that \cite{Bours2014} reported new 22~precision measurements of the HU
Aqr spanning essentially the same time period, as data gathered in our paper.
Our work extends the results of \cite{Bours2014}, since we focus on the
detailed analysis of the planetary models. We prove that the source of strong
dynamical instability in the HU Aqr planetary systems are similar semi-major
axes, placing putative companions of the binary in a~region of the 1:1~MMR and
3:2~MMR, combined with large and unconstrained masses. Moreover, we found that
planets in configurations consistent with observations are in anti-phase with
``planets'' in stable systems. 

In this paper, we gathered the new measurements in the originally submitted
manuscript and the new observations published by \cite{Bours2014}. We verified
that the results obtained without and with their measurements, respectively,
differ only up to quantitative sense (i.e., the best-fitting parameters are
slightly altered). \corr{With the aim of publishing the most up-to date
results possible}, we re-analysed all available timing data of the HU Aqr
eclipses, which are collected in Appendix.

This paper is structured as follows. Section \ref{sec:observations} presents
{22 new observations} of HU Aqr carried out between 2011 and 2014. This set
comprises of light-curves gathered with the OPTIMA photo-polarimeter hosted by
the Skinakas Observatory (Crete, Greece), as well as the most recent data
taken with {2-dim detectors operated in three different observatories, see
Tabs.~\ref{tab:tab1a}--\ref{tab:tab1b} for details}. Section \ref{sec:models}
is devoted to the Keplerian and Newtonian (O-C) models. We compare the results
from the kinematic and self consistent $N$-body models of the LTT effect. Even
more arguments are given against the kinematic model of separate \corr{LTT} orbits
\citep{Irwin1952}, which is common in the recent literature
\citep[e.g.][]{Lee2014,Hinse2014d,Almeida2013,Hinse2012,Beuermann2012,Qian2011}.
We discuss the dynamical stability of Newtonian solutions in section
\ref{sec:stability} and we conclude that at present, coplanar 2-planet and
3-planet models with direct orbits are unlikely to explain the recent (O-C)
data of HU Aqr. We show that stable 3-planet systems with highly inclined
orbits are possible. We discuss the results in section~4, estimating the (O-C)
amplitude due to the modified Applegate mechanism, and we propose independent
astrometric and imaging observations to verify the LTT hypothesis. Appendix
contains a compilation of the data set used in this paper. 
%
%
\section{New photometry of HU Aqr}
\label{sec:observations}
%
\subsection{Observations with OPTIMA in 2011 and 2012}
%
There are 22 unpublished and new mid-egress times listed in
Tab.~\ref{tab:tab2}. Among them, 15 data points were obtained with the high
time resolution photo-polarimeter OPTIMA\footnote{\tt
http://www.mpe.mpg.de/OPTIMA} \citep{Straubmeier2001, Kanbach2008,
Stefanescu2011}. OPTIMA was initially designed for optical pulsar studies,
however it is not limited to this subject only. Examples of results obtained
with OPTIMA include e.g. pulsars \citep{Slowikowska2009}, \corr{the} first
optical magnetar \citep{Stefanescu2008}, intermediate polars
\citep{Nasiroglu2012}, polars \citep{Slowikowska2013} and a black hole
candidate with optical variability \citep{Kanbach2001}.

\begin{figure*}
\centerline{
\vbox{
\hbox{\includegraphics[ width=1.0\textwidth]{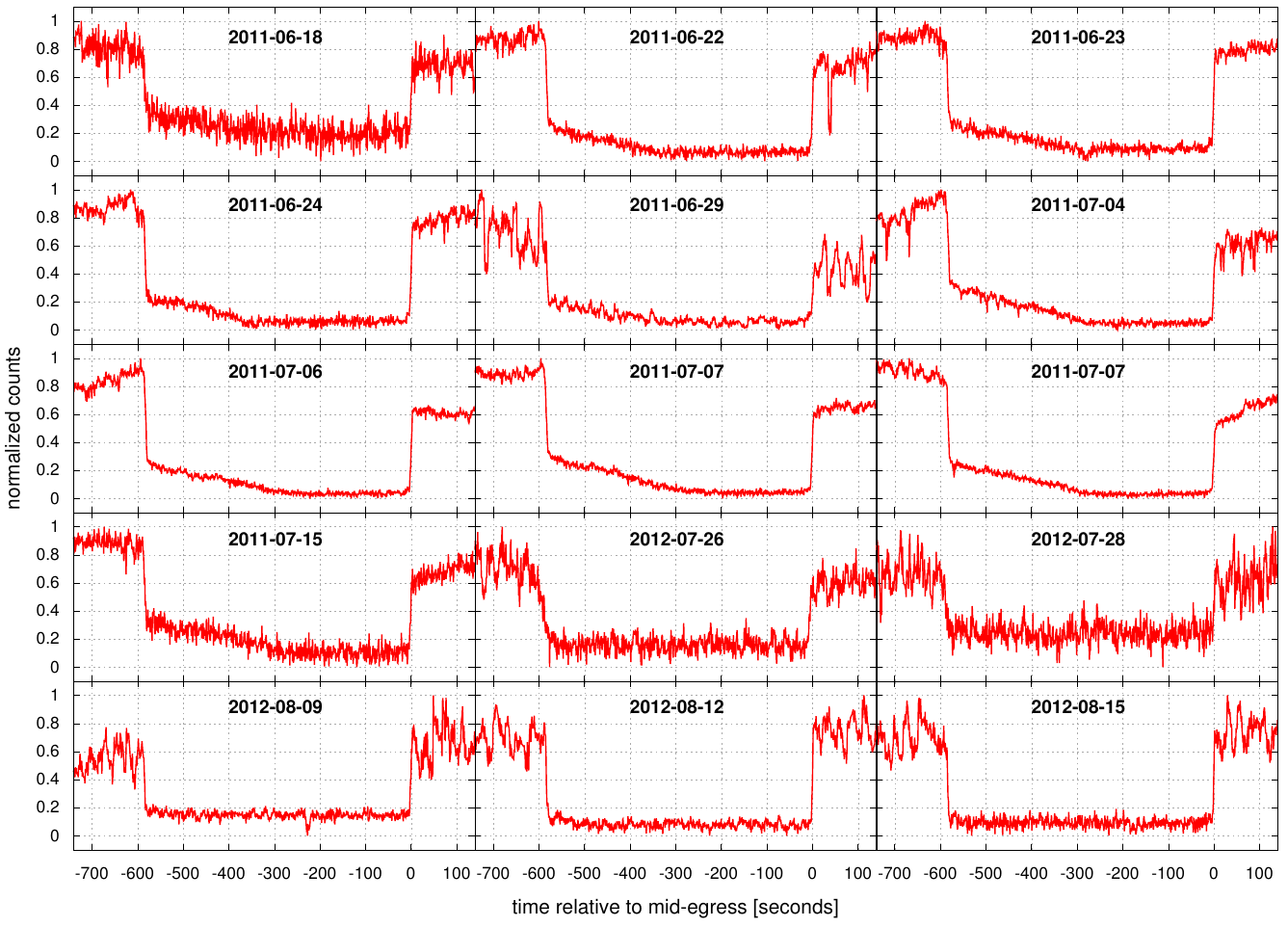}}
}
}
\caption{
{
The normalised light curves of HU Aqr obtained with OPTIMA in 2011 and 2012
(OPT-SKO). Time at the $x$-axis is relative the mid-egress moment, in accord
with Table~\ref{tab:tab2}.
}
}
\label{fig:fig1a}
\end{figure*}

Similar to previous years, we have conducted several observations of HU~Aqr
during our OPTIMA observing campaigns at the Skinakas Observatory (SKO) in
2011 and 2012 \citep{Gozdziewski2012, Slowikowska2013}. Obtained light curves
are shown in Fig.~\ref{fig:fig1a}. Using fibre-fed single photon counters,
OPTIMA is capable of recording single optical photons with an internal
accuracy of 5~nanoseconds. The absolute timing accuracy of the GPS signal is
of the order of $\sim20$--$40~\rm{}ns$. For the purpose of this paper, we bin
the {\sc OPTIMA} counts for 1~second time resolution. This is sufficient to
determine the mid-egress moment very accurately, with formal sub-second time
precision. {(We carefully checked that binning with interval of 1--3 seconds
does not change the results, hence we choose the 1~second bins to obtain
denser sampling of the light curves)}. Technical information about the
observations is gathered in Table~\ref{tab:tab1a}.

\begin{table}
\centering
\caption{
{
HU~Aqr observations with OPTIMA photo-polarimeter in 2011 and 2012
with the 1.3-m telescope at the Skinakas observatory (Crete, Greece)
in white light. Dates are given for the time of the mid-egress times.
}
}
\begin{tabular}{lccc}
\hline
\label{tab:tab1a}
Cycle & Date & Airmass & Moon Phase \\
 & & & [\%] \\
\hline
76348 & 2011-06-18 & 1.5 -- 1.7 & 86 \\
76394 & 2011-06-22 & 1.5 -- 1.6 & 52 \\
76395 & 2011-06-23 & 1.5 -- 1.6 & 52 \\ 
76406 & 2011-06-24 & 1.8 -- 1.5 & 45 \\
76464 & 2011-06-29 & 1.6 -- 1.5 & 6 \\
76532 & 2011-07-04 & 1.8 -- 1.5 & 47 \\
76555 & 2011-07-06 & 1.9 -- 1.5 & 35 \\
76556 & 2011-07-07 & 1.9 -- 1.5 & 35 \\ 
76567 & 2011-07-07 & 1.55 -- 1.6 & 47 \\
76648 & 2011-07-15 & 1.7 -- 1.55 & 100 \\
81001 & 2012-07-26 & 1.55 -- 1.48 & 57 \\
81013 & 2012-07-28 & 1.5 -- 2.2 & 68 \\
81162 & 2012-08-09 & 1.52 -- 1.47 & 48 \\
81186 & 2012-08-12 & 1.74 -- 1.84 & 21 \\
81231 & 2012-08-15 & 1.52 -- 1.55 & 3 \\
\hline\\
\end{tabular}
\end{table}

\begin{figure}
\centerline{
\vbox{
\hbox{\includegraphics[ width=0.49\textwidth]{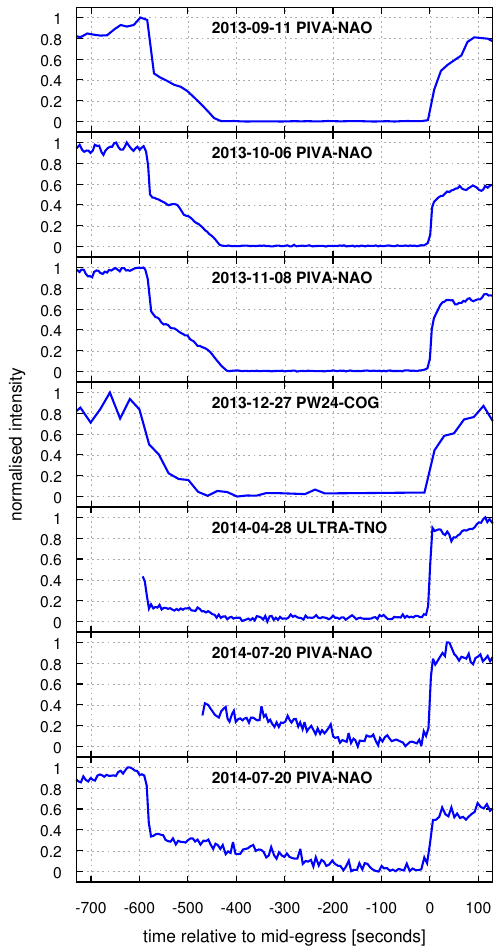}}
}
}
\caption{
The light curves of HU Aqr obtained with various CCD instruments: with the
2--meter Rozhen telescope (PIVA-NAO) in 2013 during the nights of September
11th, October 6th and November 8th and two eclipses obtained in 2014, July
20th; from PW24-COG (2013, December 27th) and from ULTRASPEC/TNO (2014, April
28th). Time at the $x$-axis is relative the mid-egress moment, in accord with
Table~\ref{tab:tab2}.
}
\label{fig:fig1b}
\end{figure}

%
\subsection{CCD-based observations in 2013 and 2014}
%
In 2013 and 2014 we performed observations at the National Astronomical
Observatory in Rozhen (NAO, Bulgaria) and at the Campus Observatory Garching
(COG, Germany). We also observed HU~Aqr with the recently inaugurated
2.4--meter Thai National Telescope (TNT) at the Thai National Observatory
(TNO, Thailand), equipped with the ULTRASPEC instrument \citep{Dhillon2014}.
Technical information about the observations is gathered in
Table~\ref{tab:tab1b}. 

Around the middle of the year 2014, we also derived a few light-curves with
the PTST 24 inch telescope at the Observatorio Astron\'omico de Mallorca (OAM,
Spain) as well as the FastCam instrument \citep{Oscoz2008} at the Teide
Observatory (TO, Spain) where we used the 1.55--meter Telescopio Carlos
S\'anchez (TCS, Spain). The most recent observations were performed on
23th--26th July 2014 at the T\"UB\.ITAK National Observatory (TUG, Turkey)
with a 1--meter robotic telescope. Unfortunately, these data have relatively
low photometric quality and are skipped in this paper.

The main telescope of the NAO Rozhen is a 2-meter Ritchey-Chretien-Coude
reflector equipped with the Princeton Instruments VersArray (PIVA): 1300 CCD
camera that has a resolution of 580$\times$550 pixels (pixel physical size of
20 microns, image scale 0.258$''$/pixel and the field of view FoV
2.5$'\times$2.36$'$). The camera is cooled down to $-110^{\circ}$C. Light
curves obtained with the 2-meter NAO telescope are shown in
Fig.~\ref{fig:fig1b}.

\begin{table}
\centering
\caption{
Parameters and conditions of HU~Aqr observations with CCD detectors. The
observatory abbreviations are provided in the text, while the filters
abbreviations stand for ``Cl'' -- clear, ``L'' -- clear, and ``\textit{g'}''
-- standard SDSS filter, respectively. "Expos." --- stands for the exposure
time. Readout time and airmass are also given. 
}
\begin{tabular}{lcccccc}
\hline
\label{tab:tab1b}
Cycle & Date & Obs./  & Expos. & Readout & Airmass \\ 
      &      & Filter & [s]    & [s]     &\\
\hline
85746 & 2013-09-11 & NAO/Cl  & 10 & $\sim$ 4    & 1.55 -- 1.70 \\
86032 & 2013-10-06 & NAO/Cl  & 3  & $\sim$ 0.5  & 1.54 -- 1.47 \\
86412 & 2013-11-08 & NAO/Cl  & 3  & $\sim$ 0.5  & 1.47 -- 1.76 \\
86976 & 2013-12-27 & COG/L   & 15 & $\sim$ 5    & 2.10 -- 2.66 \\
88383 & 2014-04-28 & TNO/g'   & 4.35  & $\sim$ 0 & 1.80 -- 1.60 \\
89339 & 2014-07-20 & NAO/Cl  & 3  & $\sim$ 2.5  & 2.30 -- 2.00 \\
89340 & 2014-07-20 & NAO/Cl  & 3 & $\sim$ 2.5 & 1.50 -- 1.45 \\
\hline\\
\end{tabular}
\end{table}

At the TNT 2.4-meter telescope we used ULTRASPEC \citep{Dhillon2014}, a
LN$_{\rm 2}$-cooled frame-transfer EMCCD with a 1024$\times$1024 active
detector area which is designed for fast, low-noise operation. Thanks to the
use of subarray windows high frame rates can be achieved, up to few 10$^2$\,Hz
\citep{Richichi2014}. Each frame is time-stamped at mid-exposure with a
dedicated GPS system. The internal timing accuracy of the system has been
tested to better than 1~ms. The observation was carried out in a standard SDSS
$g'$ filter. The resulting light curve is shown in Fig.~\ref{fig:fig1b}.
%
%
\subsection{Data analysis and timing accuracy}
%
The CCD data were reduced with the IRAF package; for NAO observations the bias
and flat field corrections were applied, while in case of COG data only dark
frames were subtracted. The TNO observations were reduced with the ULTRACAM
data reduction pipeline v9.12\footnote{\tt
http://deneb.astro.warwick.ac.uk/\\phsaap/software/ultracam/html/}, a
dedicated software for calibration and aperture photometry analysis of the
data gathered by ULTRACAM and ULTRASPEC instruments. In all cases special care
of the timing accuracy was taken. In the case of OPTIMA timing is achieved by
using a GPS receiver. In the case of NAO, the system time is synchronised
automatically every 10~min with a GPS receiver and information for corrections
is saved in a log file. Additionally, it was later controlled through NTP
server \texttt{http://time.nist.gov} and the difference was always smaller
than 0.2 sec. Time is updated every 15~min on COG at \texttt{ntp2.mpe.mpg.de}
via SNTP (MPE). Data were time stamped as JD~UTC, and for the CCD data the
mid-exposure times were taken. 

To derive the mid-egress moment, the sigmoid function representing the
photometric flux:
\begin{equation}
I(t) = a_1 + \frac{a_2 - a_1}{1.0 - \exp([t_0 - t]/\Delta t)}, 
\end{equation}
(where $a_1,a_2,\Delta t$ are parameters describing the sigmoid shape), was
fitted to selected light curves around the mid-egress moment $t_0$, within
some range of time $t$. \hide{Due to the flux variability, we carefully
selected the time--range around the mid-egress, to obtain a possibly robust
determination of its moment.} This procedure is described in Sect.~4.2 of
\citet{Gozdziewski2012}. Then the derived UTC mid-egress moments were
converted to the BJD (Barycentric Dynamical Time), using the ICRS sky
coordinates of HU~Aqr and the geodetic coordinates of each given observatory,
with the help of a numerical procedure developed by \cite{Eastman2010}. {For
all mid-egress data, we adopted the formal $1\sigma$ error of the parameter
$t_0$.} The mid--egress times obtained with all mentioned telescopes and
instruments are listed in Table~\ref{tab:tab2}.

\begin{table}
\centering
\caption{
New HU~Aqr BJD mid--egress times on the basis of light curves collected with
the OPTIMA photometer operated at the Skinakas Observatory (OPT--SKO), with
2--meter telescope at the National Astronomical Observatory (PIVA--NAO),   
with PW24 at the Campus Observatory Garching (PW24--COG), as well as with the
ULTRASPEC camera at 2.4--meter Thai National Telescope (ULTRA-TNT).
}
\begin{tabular}{lccc}
\hline
\label{tab:tab2}
Cycle $L$ & BJD & Error [day | sec] & Instrument \\
\hline
76348   & 2455731.4841422   & 0.0000054 \  |\ \ 0.47  & OPT--SKO\\
76394   & 2455735.4778822   & 0.0000025 \  |\ \ 0.22  & OPT--SKO\\
76395   & 2455735.5646948   & 0.0000023 \  |\ \ 0.20  & OPT--SKO\\
76406   & 2455736.5197228   & 0.0000019 \  |\ \ 0.16  & OPT--SKO\\
76464   & 2455741.5552777   & 0.0000093 \  |\ \ 0.80  & OPT--SKO\\
76532   & 2455747.4590878   & 0.0000029 \  |\ \ 0.25  & OPT--SKO\\
76555   & 2455749.4559585   & 0.0000017 \  |\ \ 0.15  & OPT--SKO\\
76556   & 2455749.5427747   & 0.0000016 \  |\ \ 0.14  & OPT--SKO\\
76567   & 2455750.4978035   & 0.0000019 \  |\ \ 0.16  & OPT--SKO\\
76648   & 2455757.5302482   & 0.0000071 \  |\ \ 0.61  & OPT--SKO\\
81001   & 2456135.4591635   & 0.0000071 \  |\ \ 0.61  & OPT--SKO\\
81013   & 2456136.5010300   & 0.0000058 \  |\ \ 0.50  & OPT--SKO\\
81162   & 2456149.4372694   & 0.0000059 \  |\ \ 0.51  & OPT--SKO\\
81186   & 2456151.5209490   & 0.0000020 \  |\ \ 0.17  & OPT--SKO\\
81231   & 2456155.4278501   & 0.0000034 \  |\ \ 0.29  & OPT--SKO\\
85746   & 2456547.4214776   & 0.0000073 \  |\ \ 0.63  & PIVA-NAO\\
86032   & 2456572.2520793   & 0.0000034 \  |\ \ 0.29  & PIVA-NAO\\
86412   & 2456605.2437774   & 0.0000027 \  |\ \ 0.24  & PIVA-NAO\\
86976   & 2456654.2104280   & 0.0000361 \  |\ \ 3.12  & PW24-COG\\
88383   & 2456776.3665044   & 0.0000017 \  |\ \ 0.14  & ULTRA-TNT\\
89339   & 2456859.3666325   & 0.0000064 \  |\ \ 0.56  & PIVA-NAO\\   
89340   & 2456859.4534517   & 0.0000103 \  |\ \ 0.89  & PIVA-NAO\\  
\hline 
\end{tabular}
\end{table}
%
%
\section{LTT models of the (O-C)}
%
\label{sec:models}
\label{subsec:model}
To revise the LTT models of \citet{Gozdziewski2012}, we use the collected
mid-egress moments published in \citet{Schwope2001}, \citet{Schwarz2009},
\citet{Gozdziewski2012}, updated by our new observations displayed in
Table~\ref{tab:tab2} {and by the recent 22 data points published by
\cite{Bours2014}. Summarising, the new 44 data points extend the set of all
previous literature data used for constraining $N$-body models in
\citet{Gozdziewski2012}}. A complete list of 215 mid-egress moments is
compiled in Appendix (Table~\ref{tab:tab5}). The zero cycle ($L=0$) epoch for
the third body models is $T_0=$ JD~2,453,504.88829400, roughly in the middle
of the observation window, and of the most accurate OPTIMA measurements. Note
that Table~\ref{tab:tab5} displays eclipse cycles counted from epoch
JD~2449102.9200026, i.e., the epoch of first observation of HU~Aqr in
\citet{Schwope1991}. Though the mid-egress measurements in \citet{Qian2011}
are included in Table~\ref{tab:tab5}, they are not used here. These data
systematically outlay by a~few seconds from more accurate OPTIMA and MONET/N
timings, spanning the same observational window \citep{Gozdziewski2012}. Note
that \cite{Bours2014} also did not include the \cite{Qian2011} observations
for the same reason.  
\begin{figure*}
\centerline{
  \includegraphics[width=0.5\textwidth]{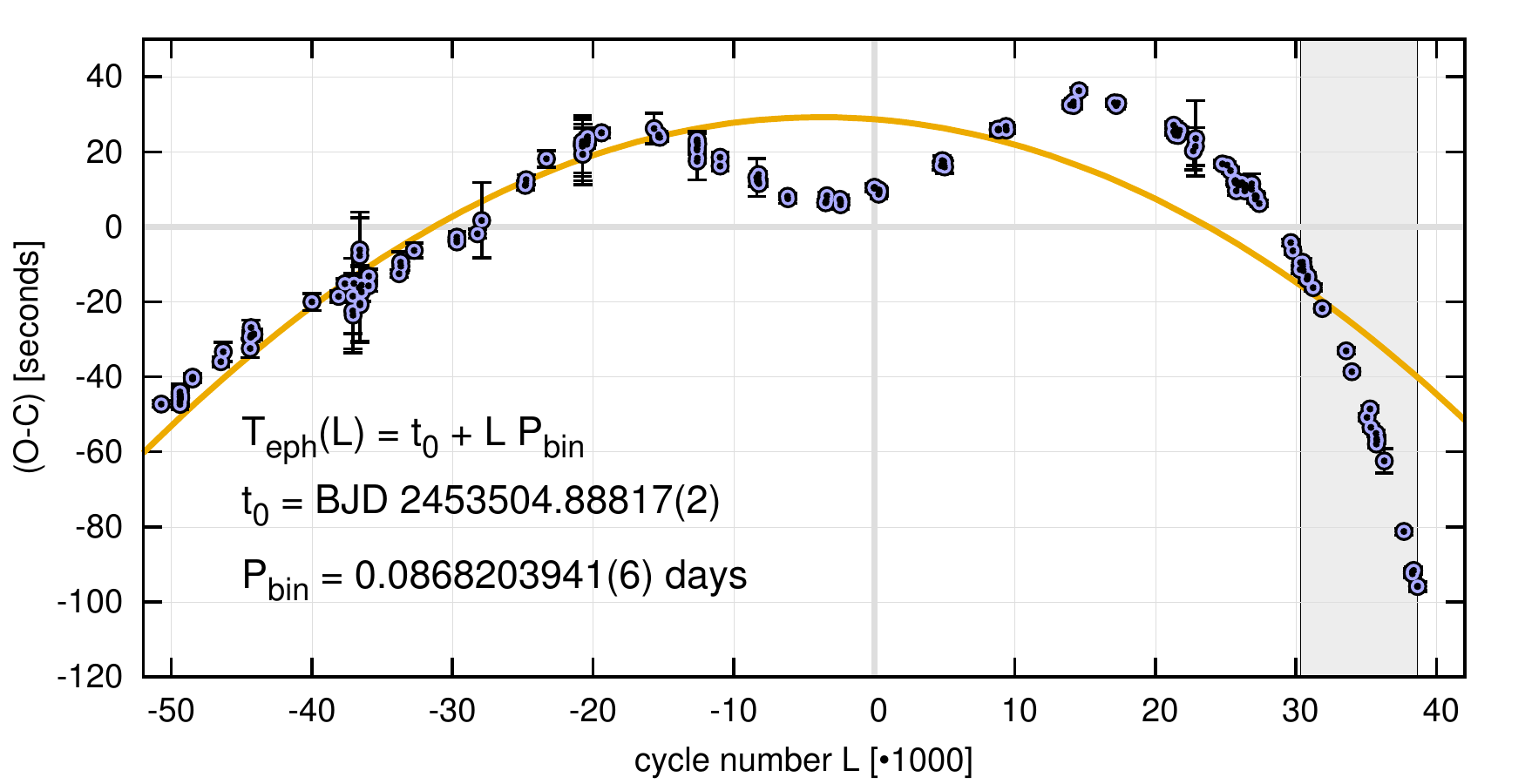}
  \includegraphics[width=0.5\textwidth]{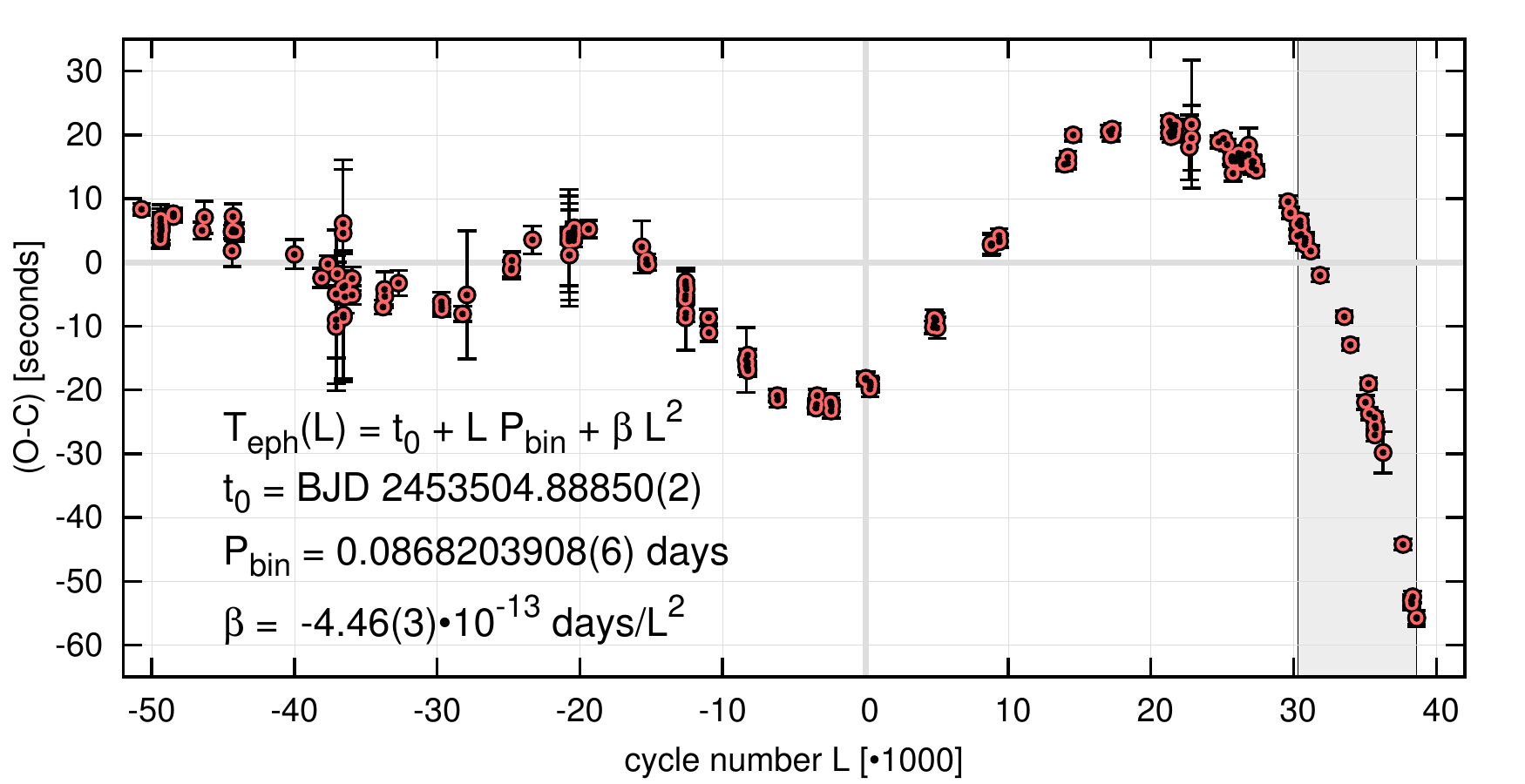}
}
\caption{
The (O-C) diagrams \corr{for mid-egress moments of HU Aqr collected in this paper
(see Appendix, Tab.~\ref{tab:tab5}), with the ``jitter'' error
correction of $\sigma_f = 0.9$~seconds}, see the text for details. Red and blue
filled circles are for the linear ({\em the left-hand panel}) and for
parabolic ({\em the right-hand panel}) ephemeris, respectively. The yellow
curve in {\em the left-hand} panel illustrates the parabolic term in
Eq.~\ref{eq:ephem}. Labels are for the fit parameters to Eq.~\ref{eq:ephem}
with uncertainties at the last significant digit, which is marked with a digit
in parenthesis. The shaded region is for the new data that show \corr{a
large} deviation from the previous predictions in the literature.
}
\label{fig:fig2}
\end{figure*}
 
With these data, we construct (O-C) diagrams (Fig.~\ref{fig:fig2}) for the
linear and parabolic ephemeris, respectively:
\begin{equation}
T_{\idm{ephem}}(L) = t_0 + L P_{\idm{bin}} + \beta L^2,
\label{eq:ephem}
\end{equation}
where $T_{\idm{ephem}}(L)$ is the time of predicted mid-egress at eclipse
cycle $L$, $t_0$ is the epoch, and $\beta$ is the derivative of the orbital
period $P_{\idm{bin}}$, in accord with \citet{Hilditch2001}.

Significant (O-C) mid-egress deviations from the linear ephemeris (blue filled
circles in Fig.~\ref{fig:fig2}, the left-hand panel) and from the parabolic
ephemeris (red filled circles in the right-hand panel) have become apparent
shortly after the end of observing season 2012 (see at the left border of
shadowed rectangles). We continued to monitor the target in September, 2013
(NAO, Rozhen). Around this epoch, the (O-C) for the parabolic ephemeris in
\citet{Gozdziewski2012} are already $\sim60$~s. Such a large magnitude was
very unexpected, and we tried to verify and confirm the NAO timing data with
other instruments. It was possible only in December, 2013 through observations
with the PW24-COG instrument (see Tab.~\ref{tab:tab1b}). Shortly thereafter,
\citet{Schwope2014} published 6~observations performed with a small 14~inch
Celestron telescope, between Oct 22 and Oct 30, and new ephemerides of the HU
Aqr eclipses, which confirmed our findings. (We do not use their observations
in this paper). Continued monitoring of the HU Aqr during the new observing
season 2014 revealed progressing decay of (O-C). Recent (O-C) reach $120$~s
for the linear ephemeris, and more than $60$~s for the parabolic ephemeris
(Fig.~\ref{fig:fig2}). The parabolic term about of $-5 \cdot
10^{-13}$~day\,$L^{-2}$ is larger than its previous estimates
\citep[e.g.][]{Gozdziewski2012,Qian2011,Schwarz2009}.
%
\subsection{Keplerian model of the LTT and optimisation algorithms}
%
\label{subsec:oc}
In \citet{Gozdziewski2012}, we revised a common kinematic (Keplerian)
formulation of the LTT effect \citep{Irwin1952} for multiple companions of the
binary. This is accomplished by expressing eclipse ephemerides w.r.t. Jacobi
coordinates with the origin at the centre of mass (CM) of the binary. Because
the binary period is shorter than the orbital periods by a factor of
$\sim10^{5}$, the binary is well approximated by a point in the CM with the
total mass of both stars; {the mass of the HU Aqr binary is
0.98~$M_{\sun}$~\citep{Schwope2011}}. Osculating orbital elements and masses
derived in this way best match the true, $N$-body initial condition of the
system with mutually interacting planets.

This ephemeris model accounting for the presence of planetary companions has
the more general form of Eq.~\ref{eq:ephem}:
\begin{equation}
T_\idm{ephem}(L) = t_0 + L P_{\idm{bin}} + \beta L^2 + \sum_p \zeta_p(t(L)),
\label{eq:epla}
\end{equation}
where the $\zeta_p(t)$ terms are for the (O-C) deviations induced
by gravitational perturbation of the CM by the third bodies ($p=1,2,\ldots$
or, in accord with the common convention $p=\rm{b},\rm{c},\ldots$):
\[
\zeta_p(t) = K_p \left[ \sin \omega_p \left( \cos E_p(t) - e_p \right) 
      + \cos \omega_p \sqrt{1-e_p^2} \sin E_p(t) \right],
\]
and $K_p, e_p,\omega_p$ are the semi-amplitude of the LTT signal,
eccentricity, and argument of the pericenter, respectively for body $p$. Its
orbital period $P_p$ and the pericenter passage $T_p$ are introduced
indirectly through the eccentric anomaly $E_p(t)$. Details are given in
\citet{Gozdziewski2012}, see also further development of this idea and
discussion in \citet{Marsh2014}. 

The initial (O-C) diagrams in Fig.~\ref{fig:fig2} are suggestive for
a~significant parabolic term, which is also often quoted in the literature.
Therefore, we optimised the general form of Eq.~\ref{eq:epla} including a
parameter $\beta$ that accounts for a secular change of the binary period
$P_{\idm{bin}}$ (i.e., the quadratic ephemeris). After extensive analysis, we
found that an alternative 2-planet model with the linear ephemeris is heavily
unconstrained with respect to a few parameters. Its formal best-fitting
solution has similar orbital periods $\sim 22$~years, and permits LTT
semi-amplitudes $K_{1,2}$ as large as $60$~min. This implies the non-physical
companion masses of $\sim 1\,$M$_\odot$ (this will be shown at the end of this
section). Moreover, the linear ephemeris model provides statistically worse
solutions in terms of an rms, as compared to the 2-planet quadratic ephemeris,
i.e., the fit model with only one free parameter more. The quadratic term,
usually interpreted as the binary period derivative \citep{Hilditch2001},
might be also caused by a~distant and massive companion with a very long
orbital period. Hence we focus mainly on the parabolic ephemeris variant of
the 2-companion hypothesis.

To optimise the ephemeris models in Eq.~\ref{eq:epla} in terms of {the reduced
$\Chi$ function}, we apply a combination of the Genetic Algorithms
\citep[GA,][]{Charbonneau1995} and the local and fast Levenberg-Marquardt
scheme, as described in \citet{Gozdziewski2012}. This hybrid algorithm (HA
from hereafter) provides an efficient and robust exploration of
multidimensional parameter space. Each run of the HA starts from the GA step,
which searches for reasonable solutions over wide ranges of model parameters.
In particular, the hypercube of parameters explored by the GA has
$K_{\idm{b,c}} \in [1,1800]$~s, $P_{\idm{b,c}} \in [2000,36000]$~days,
$e_{\idm{b,c}} \in [0,0.99]$, and all angles permitted in their full ranges.
All GA--derived sets of solutions are then refined with the
Levenberg-Marquardt method. This procedure repeated many times makes it
possible to gather large statistics of solutions which are consistent with the
observations. (HA may be also used to optimise the $N$-body models). 

\subsection{Correction of the timing uncertainties}
At the preliminary stage, we found that in spite of a wider observational
window than in \citet{Gozdziewski2012}, the HA converged to non-unique
solutions with marginally different $\Chi$. The best-fitting solutions exhibit
$\Chi\sim 9$ on the raw data in Table~\ref{tab:tab5}. Such a large value may
indicate an incorrect fit model. However, recalling the quasi-periodic
character of the (O-C) shown in Fig.~\ref{fig:fig2} and astrophysical
arguments, we assumed that the LTT model may be valid. Therefore, the second
possibility are underestimated uncertainties of the measurements
\citep{Bevington}. This makes it difficult to derive proper uncertainties of
the fit parameters. Hence, assuming that the mid-egress errors are normally
distributed, we examined the uncertainty correction, similar to the so called
stellar ``jitter'', which is a well known factor that must be accounted for in
the RV technique \citep[e.g.][]{Butler2003,Wright2005}. In simple settings, a
one-parameter jitter uncertainty describes intrinsic RV variability that is
caused by the stellar chromosphere. Here, we consider a similar correction to
the mid-egress timing error, which accounts, for instance, for the unmodeled
geometry of eclipses, or observational circumstances and instrumental errors,
e.g. additional elements of emission or absorption of light in the binary
system. We add such a'priori unspecified term $\sigma_f$ in quadrature to
uncertainties of individual observations $\sigma_i$, through $\sigma_i^2
\rightarrow \sigma_i^2 + \sigma_f^2$, where $i=1,\ldots,N_{\idm{obs}}$, and
$N_{\idm{obs}}$ is the total number of measurements. The $\sigma_f$ term is
then the free parameter of the fit model. 

The introduction of the $\sigma_f$ factor is supported by arguments of
\cite{Schwope2014}. They argue that formal uncertainties of the midegress
moments below $\sim 1$ second are much smaller than the finite physical size
of the accretion spot. The $X$-ray emitting region is $\sim450$~km across wide
\citep{Schwope2001}, and is wider in the optical domain since the egress phase
lasts for $\sim7$-$8$~seconds. Moreover, the accretion area migrates over the
WD surface, hence the geometry of eclipses is changing, and introducing a
short-term component of the (O-C), which is estimated for $1$--$2$ seconds
\citep{Schwarz2009}. Small timing errors are possible for densely sampled
light curves, and are reported, for instance, by \cite{Schwope2001} for {\sc
ROSAT} (quoted errors for cycles 2212--2225 are $\sim0.13$--$0.24$~seconds),
by \cite{Gozdziewski2012} for {\sc OPTIMA} data (quoted errors are on the
level of $\sim 0.1$--$0.8$~seconds), by \cite{Schwarz2009} for {\sc
UTRACAM-VLT} (errors $\sim 0.5$~seconds), and, very recently by
\cite{Bours2014} for the {\sc ULTRASPEC-TNT} instrument (the quoted errors are
as small as 0.02--0.06 seconds). 
 
To determine $\sigma_f$ for selected solutions found with the HA, we optimised
the maximum likelihood function ${\cal L}$:
\begin{equation}
 \log {\cal L} =  
-\frac{1}{2} \sum_i \frac{{\mbox{(O-C)}}_i^2}{(\sigma_i^2+\sigma_f^2)}
- \sum_i \log \sqrt{{\sigma_i^2+\sigma_f^2}} -N \log{\sqrt{2\pi}},
\label{eq:Lfun}
\end{equation}
where $(\mbox{O-C})_i$ is the (O-C) deviation of the mid-egress at given
eclipse cycle $L_i$ (Eq.~\ref{eq:epla}). This form of ${\cal L}$ is similar
to that one introduced by \cite{Baluev2008} for the RV data.

Furthermore, to analyse the parameter correlations in detail, we
performed the Markov Chain Monte Carlo (MCMC) analysis of selected
best-fitting solutions. The posterior probability distribution of model
parameters $\tv{\theta}$ given the data set ${\cal D}$ is defined through
\[
   p(\tv{\theta}|{\cal D}) \propto p(\tv{\theta}) p({\cal D}|\tv{\theta}),
\]
where $p(\tv{\theta})$ is the prior, and the sampling data distribution
$p({\cal D}|\tv{\theta}) \equiv \log{\cal L}(\tv{\theta},{\cal D})$. We
defined the parameter priors as uniform or uniform improper through placing
only natural, physical and geometric limits on the parameters, i.e.,
$K_{\idm{p}}>0$, $P_{\idm{p}}>0$, $e_{\idm{p}} \in [0,1)$, $\omega_{\idm{p}}
\in [0,2\pi]$, $T_{\idm{p}}>0$, $P_{\idm{bin}}>0$, $\beta<0$, and
$\sigma_f>0$, where $\mbox{p}$ is the planet index,
$\mbox{p}=\mbox{b},\mbox{c},\mbox{d}$, and so on. These computationally simple
priors are justified here, since we analyse well localised solutions.
Moreover, we verified that the modified Jeffreys prior for parameter $\theta
\equiv \sigma_f$, as introduced by \cite{Gregory2005},
\[
p({\theta}) = \frac{1}{(\theta +\theta_{\idm{min}})} 
\frac{1}{\log [(\theta_{\idm{min}} + \theta_{\idm{max}})/\theta_{\idm{min}}]},
\]
(where $\theta_{\idm{min}}$ and $\theta_{\idm{max}}$ are scaling constants,
fixed for $\sigma_f$ as equal to 0.01~s and 10~s, respectively), does not
change the results. Similar priors have been defined for the $N$-body models,
with planetary masses $m_{\idm{p}}>0$, semi-major axes $a_{\idm{p}}>0$,
eccentricities $e_{\idm{p}} \in [0,1)$, and all Keplerian angles $\in
[0,2\pi]$.

To perform the optimisation of $\log{\cal L}$ and the MCMC analysis, we
prepared {\sc Python} interfaces to model functions written originally in {\sc
Fortran} and we used publicly available, excellent {\tt emcee} code of the
affine-invariant ensemble sampler for MCMC proposed by \cite{goodman2010},
kindly provided by \cite{foreman2014}. See also the recent paper by
\citet{Marsh2014} for a consistent application of the MCMC optimisation and
this code to the analysis of the (O-C) diagrams.

As shown below, all LTT models of HU Aqr studied in this paper suffer from
strong parameter correlations and are not unique (the posteriors are
multi-modal). In such a case, an efficient application of the MCMC to explore
the whole parameter space is very difficult \citep{foreman2014,Marsh2014}. In
this sense, we found that the hybrid algorithm and MCMC are complementary. The
MCMC method is a~great tool to analyse properties of the best fitting
configuration found with the hybrid algorithm, and is very useful to derive
realistic uncertainties of the parameters.

The results for the best-fitting Keplerian 2-planet quadratic ephemeris model
(Table~\ref{tab:tab3}) are illustrated in the form of 1-dim and 2-dim
projections of the posterior probability distributions. The best-fitting
models obtained in this way have slightly altered parameters, as compared to
their values derived through standard minimisation of $\Chi$. We {obtained
$\sigma_f \sim 0.9$--$1.5$~seconds for different LTT variants (i.e., 2-planet
and 3-planet configurations with the linear and parabolic ephemeris) and the
most recent data set}. After applying the correction term, we found that
$\Chi$ is $\sim 1$, and the rms remains unaltered, as expected. Furthermore,
we ran the HA optimisation code on all mid-egress data in
Table~\ref{tab:tab5}, with uncertainties added in quadrature to a constant
value of $\sigma_f=0.9$~s. {Yet individual best-fitting models were recomputed
with $\sigma_f$ as a free parameter.} The $\sigma_f$ correction is most
significant for {\sc ROSAT}, {\sc OPTIMA} and {\sc ULTRASPEC} measurements
with formal sub-second accuracy. (Note, that mid-egress timing data in
Table~\ref{tab:tab5} do not include this term in the error column). 

It may be easily overlooked that the error correction has a~significant
influence on the parabolic ephemeris itself, see the right-hand panel in
Fig.~\ref{fig:fig2}, which shows the (O-C) diagram computed for data errors
corrected with $\sigma_f=0.9$~seconds. Without this correction, $\beta \simeq
-5.4 \cdot 10^{-13}~\mbox{day}\cdot L^{-2}$, and the residual (O-C) signal to
be explained by the LTT model is clearly modified, when compared to the raw
timing data.
%
%
\subsection{Quadratic ephemeris, kinematic 2-planet model}
\begin{figure*}
\centerline{
 \hbox{\includegraphics[ width=0.44\textwidth]{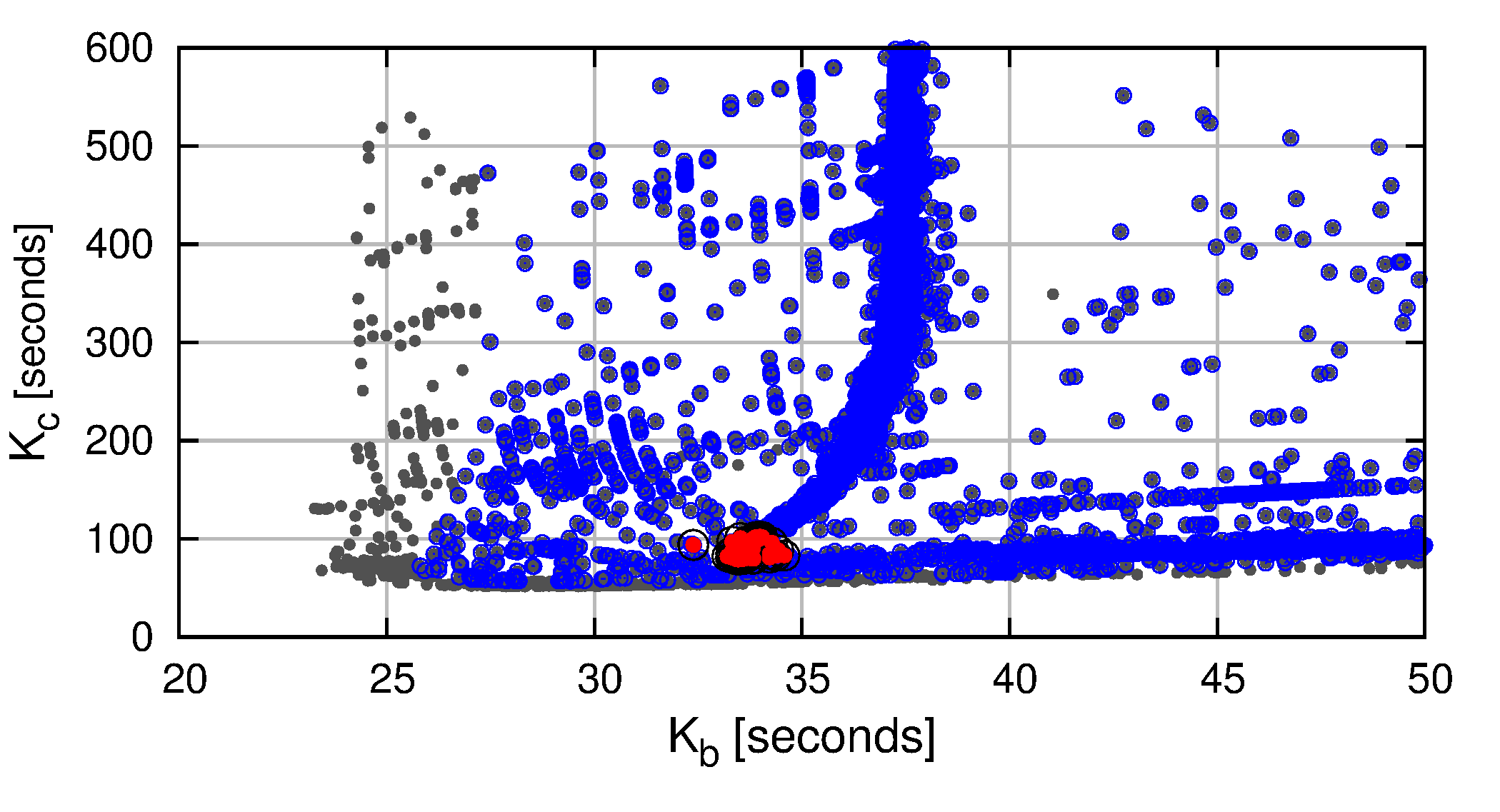}}
 \hbox{\includegraphics[ width=0.44\textwidth]{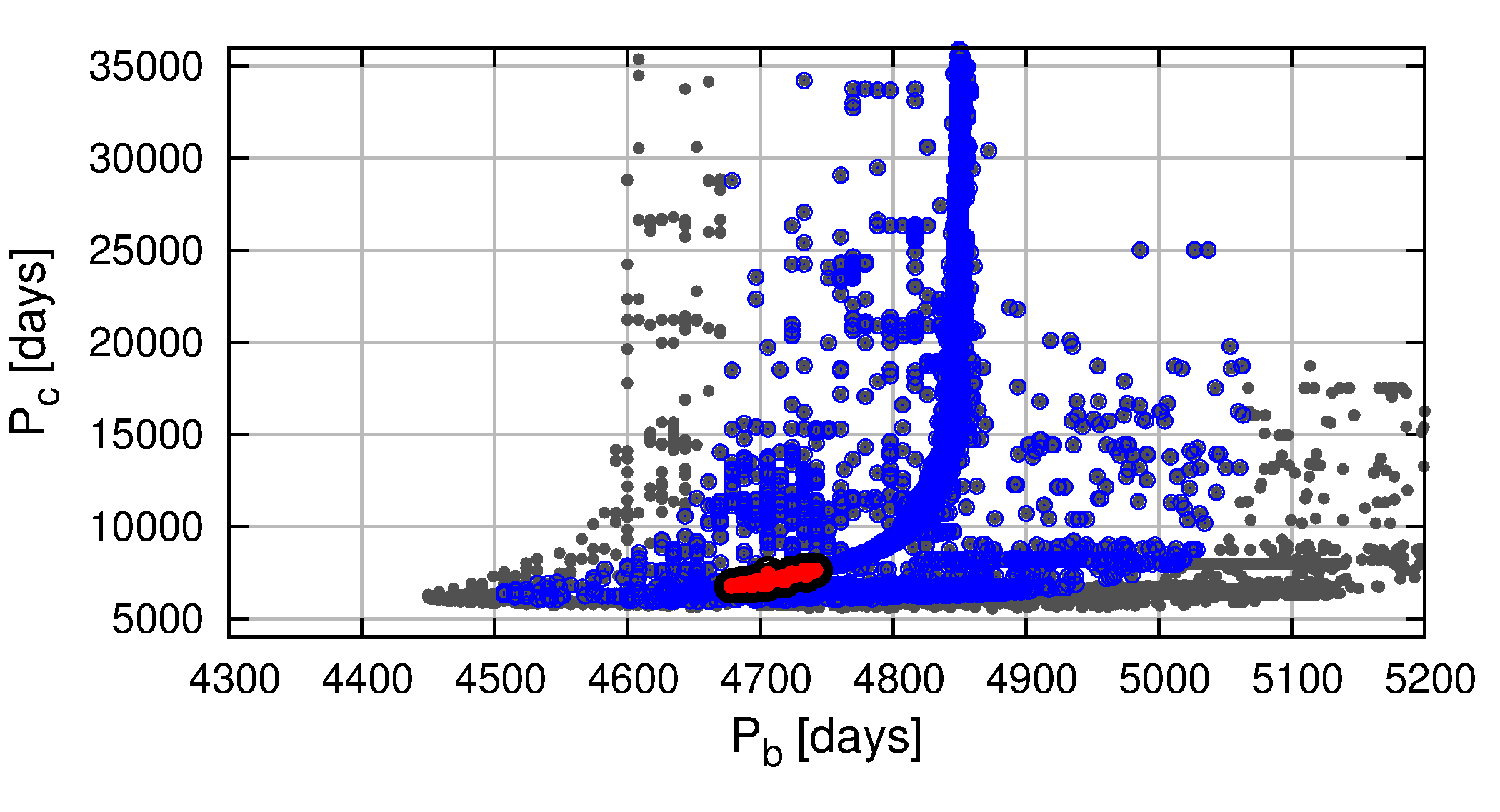}}
}
\centerline{
 \hbox{\includegraphics[ width=0.44\textwidth]{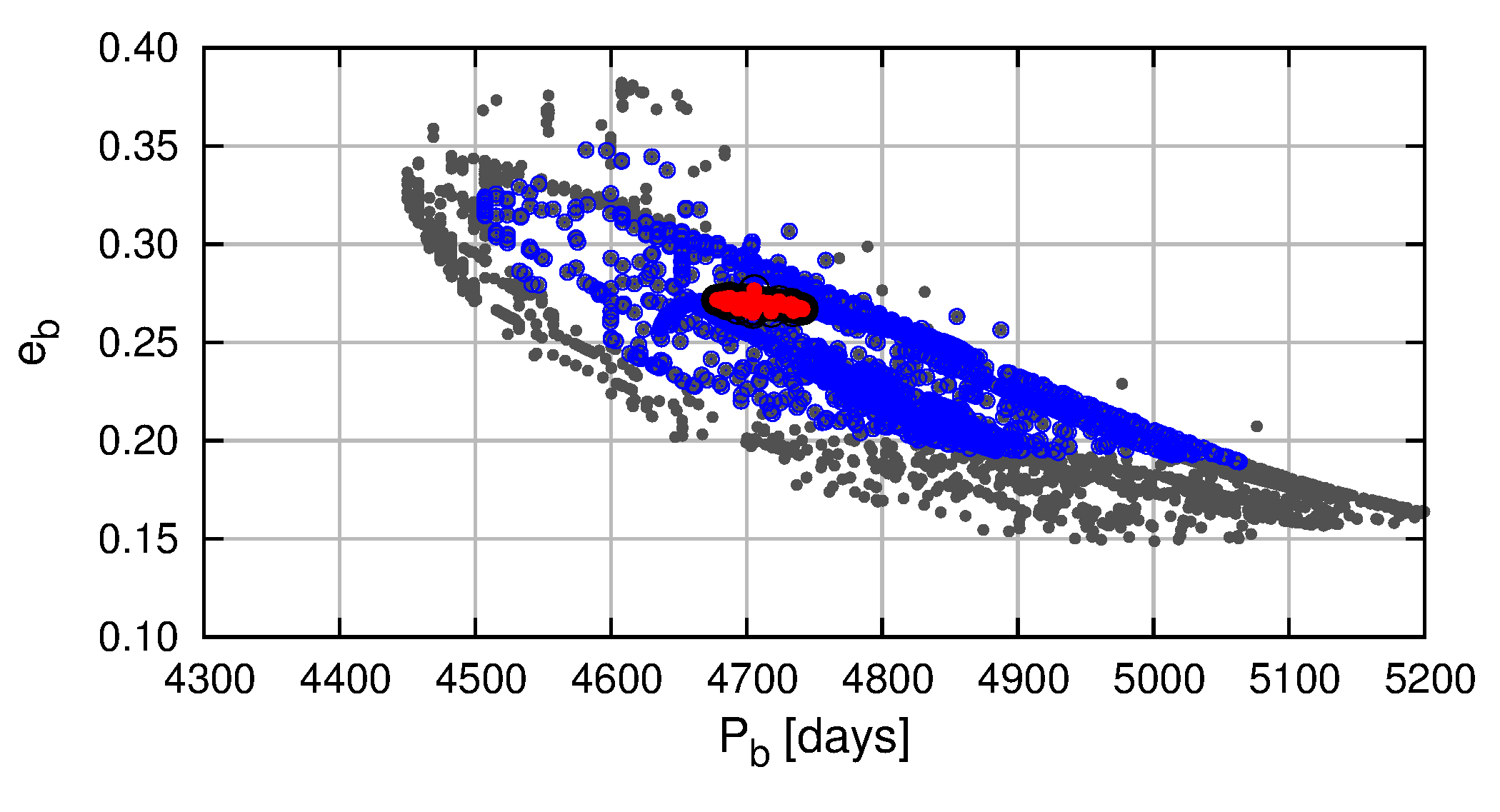}}
 \hbox{\includegraphics[ width=0.44\textwidth]{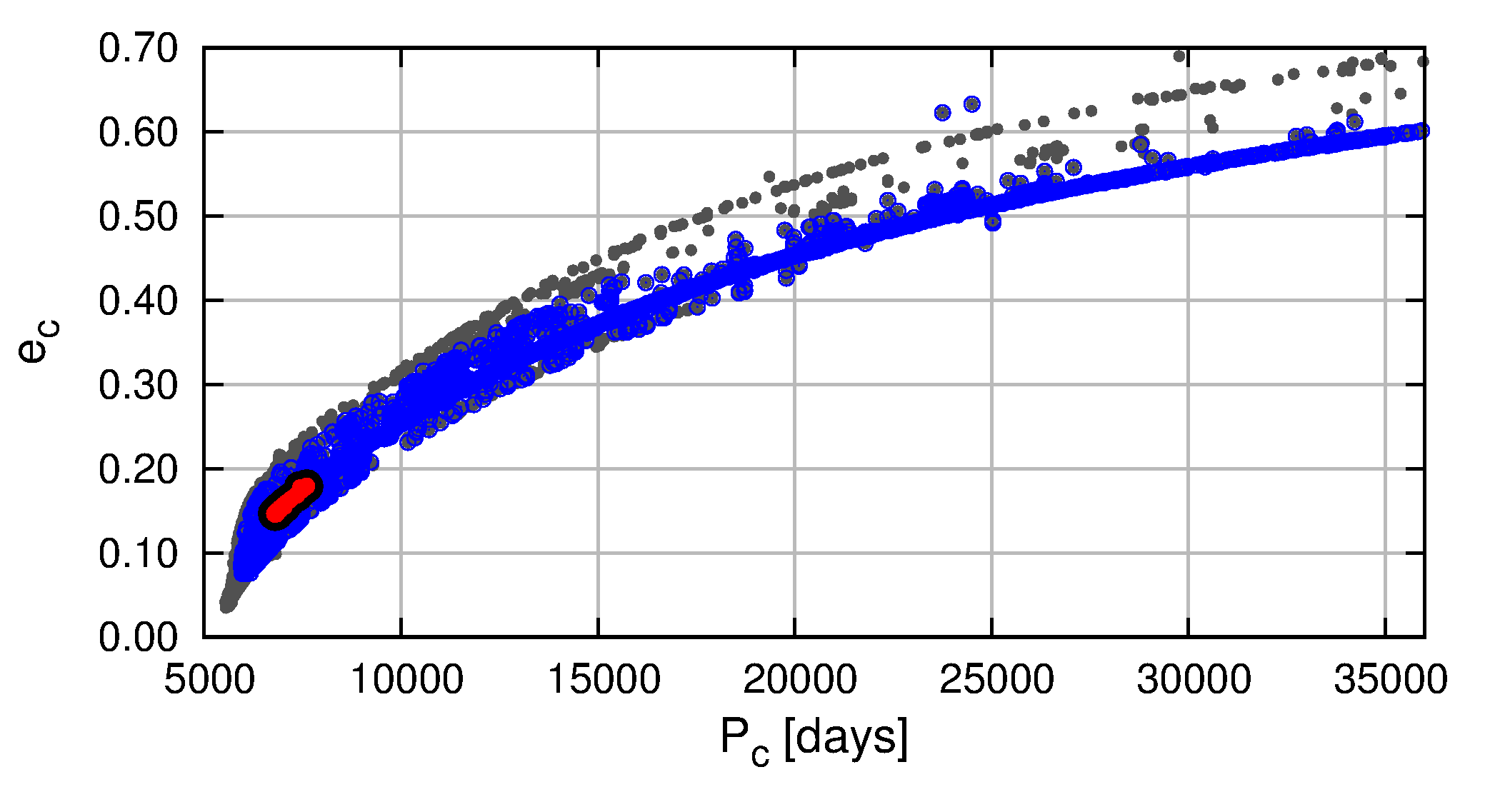}}
}
\centerline{
 \hbox{\includegraphics[ width=0.44\textwidth]{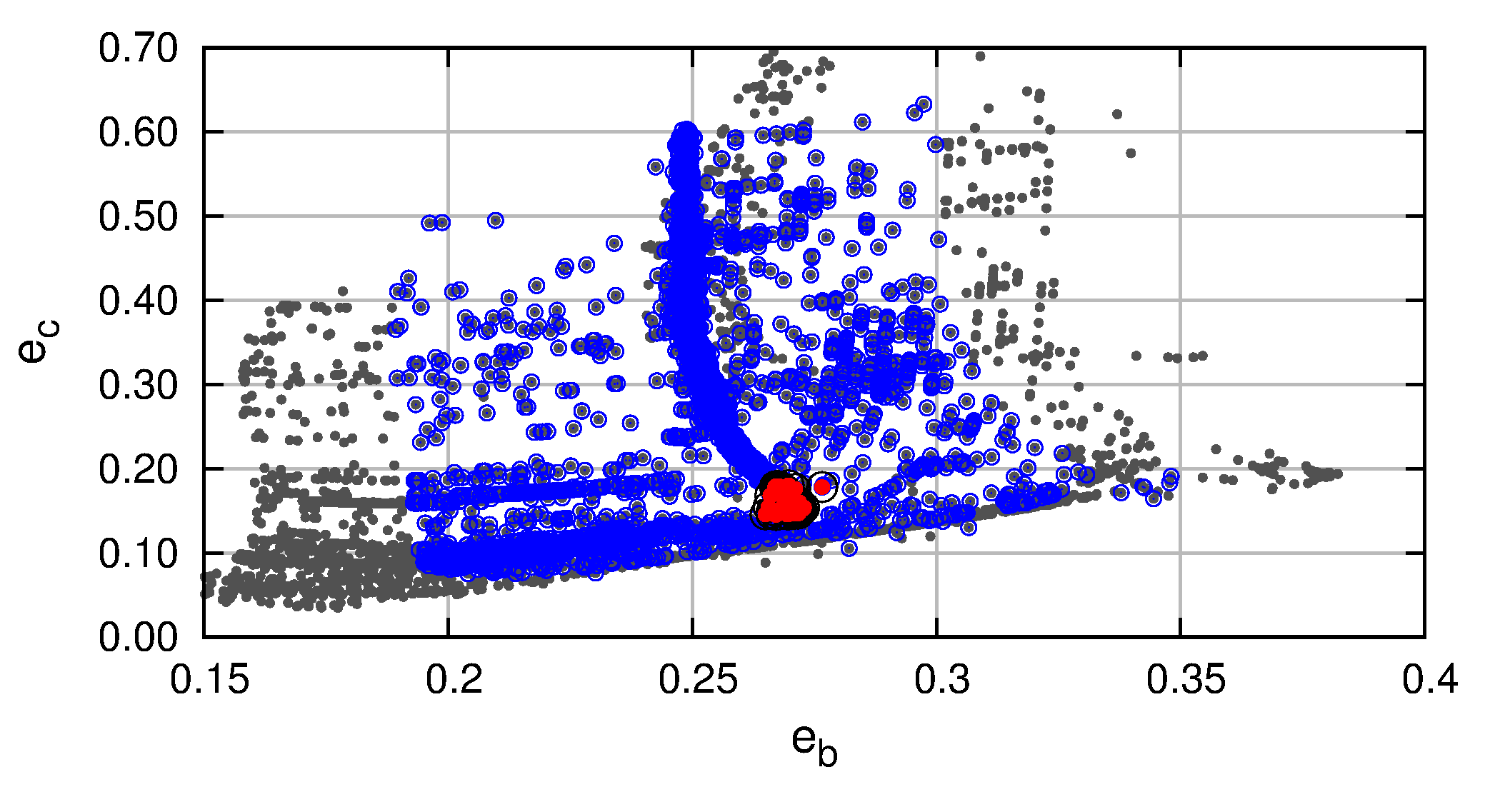}}
 \hbox{\includegraphics[ width=0.44\textwidth]{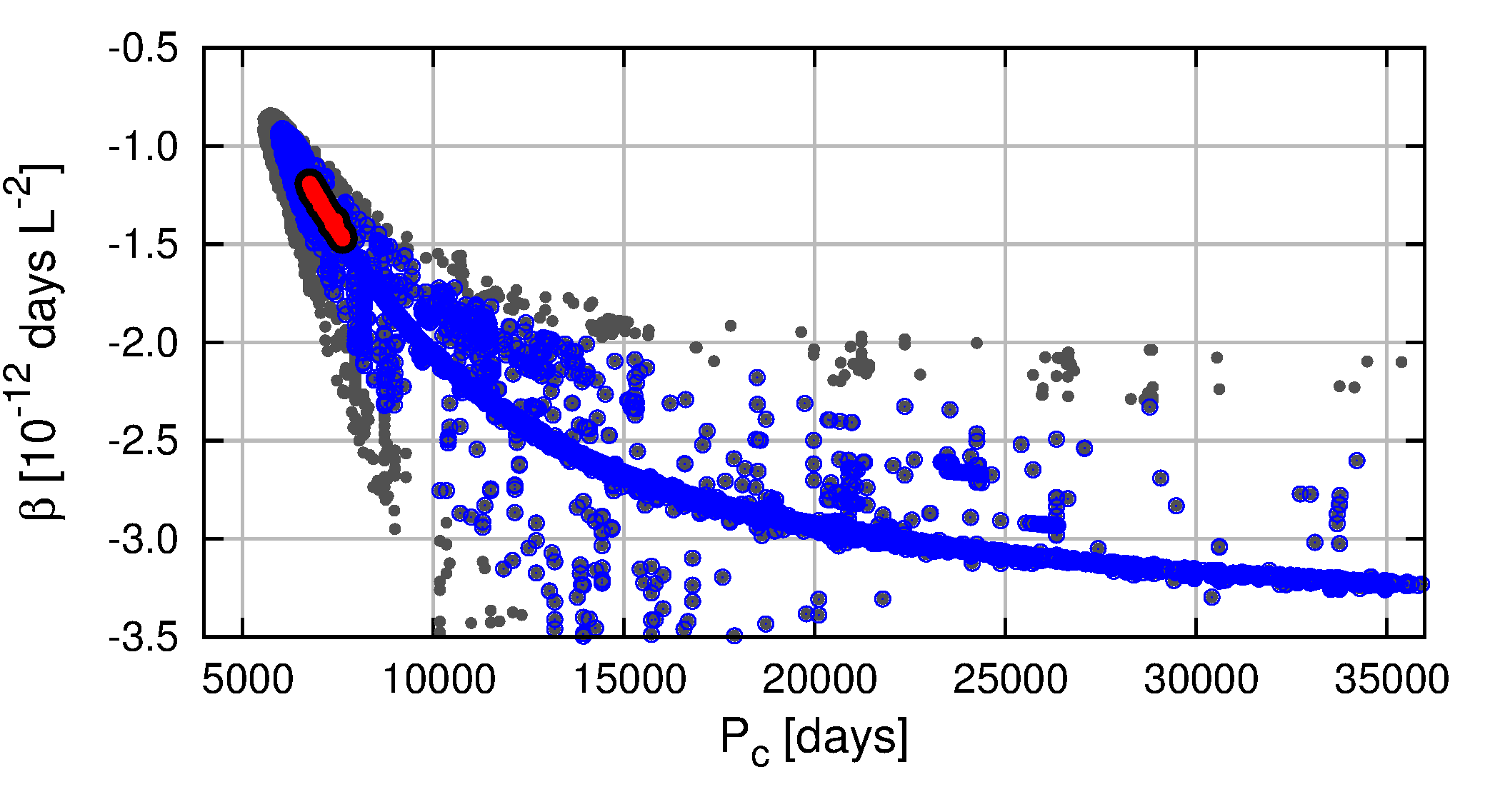}}
}
\caption{
Best-fitting 2-planet Jacobian solutions to the quadratic ephemeris projected
onto selected parameter planes. Red circles are for models with $\Chi<0.99$
(marginally better than $\Chi = 0.986$ of the best-fitting model JQ see
Tab.~\ref{tab:tab3}), blue open circles and grey filled circles are for
$\Chi<1.04$ and $\Chi<1.08$, respectively (roughly $1\sigma$ for
$3\sigma$--confidence intervals of the best-fitting solution).
}
\label{fig:fig3}
\end{figure*}

\begin{figure*}
\centerline{
 \hbox{\includegraphics[ width=0.48\textwidth]{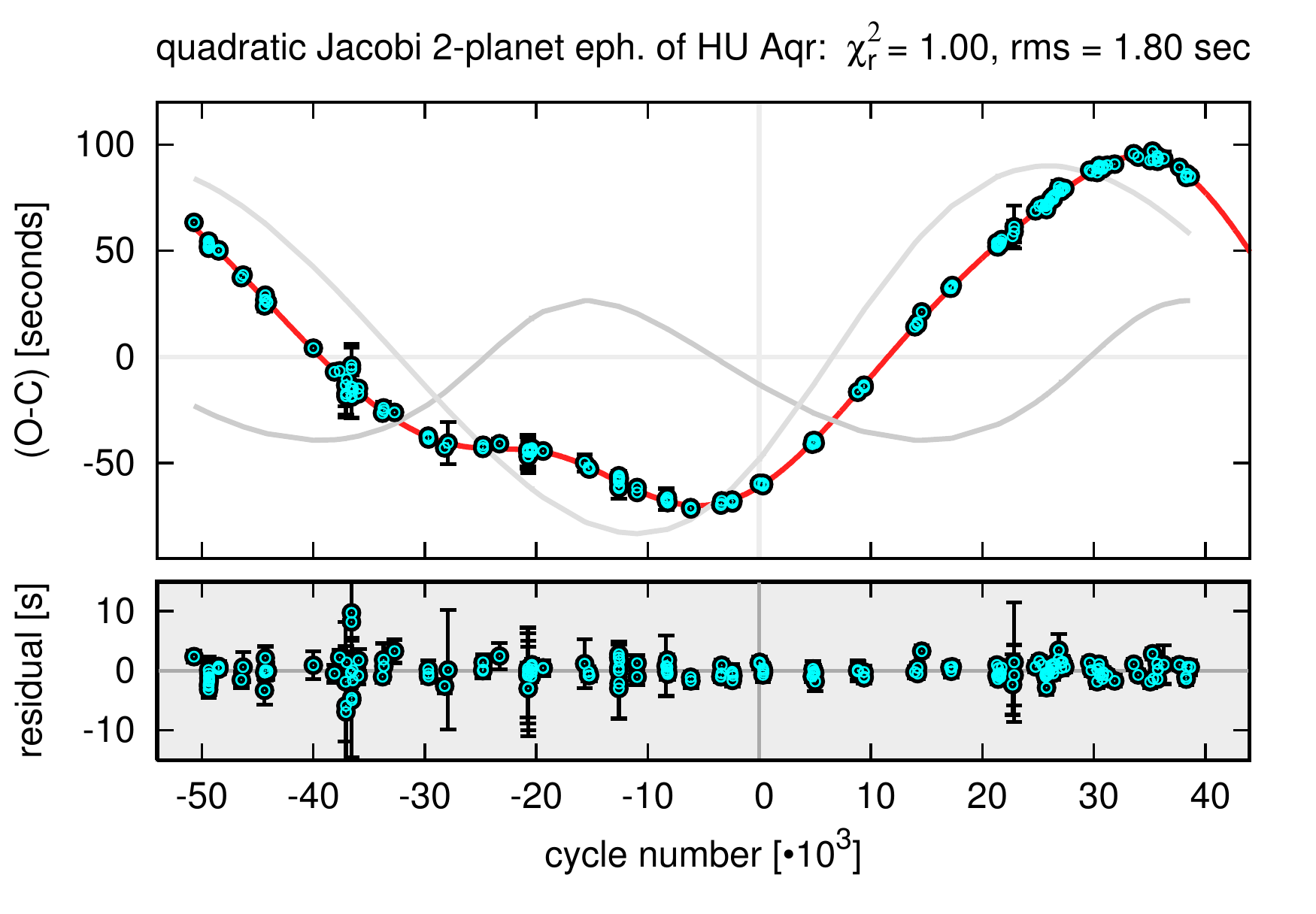}}
 \hbox{\includegraphics[ width=0.48\textwidth]{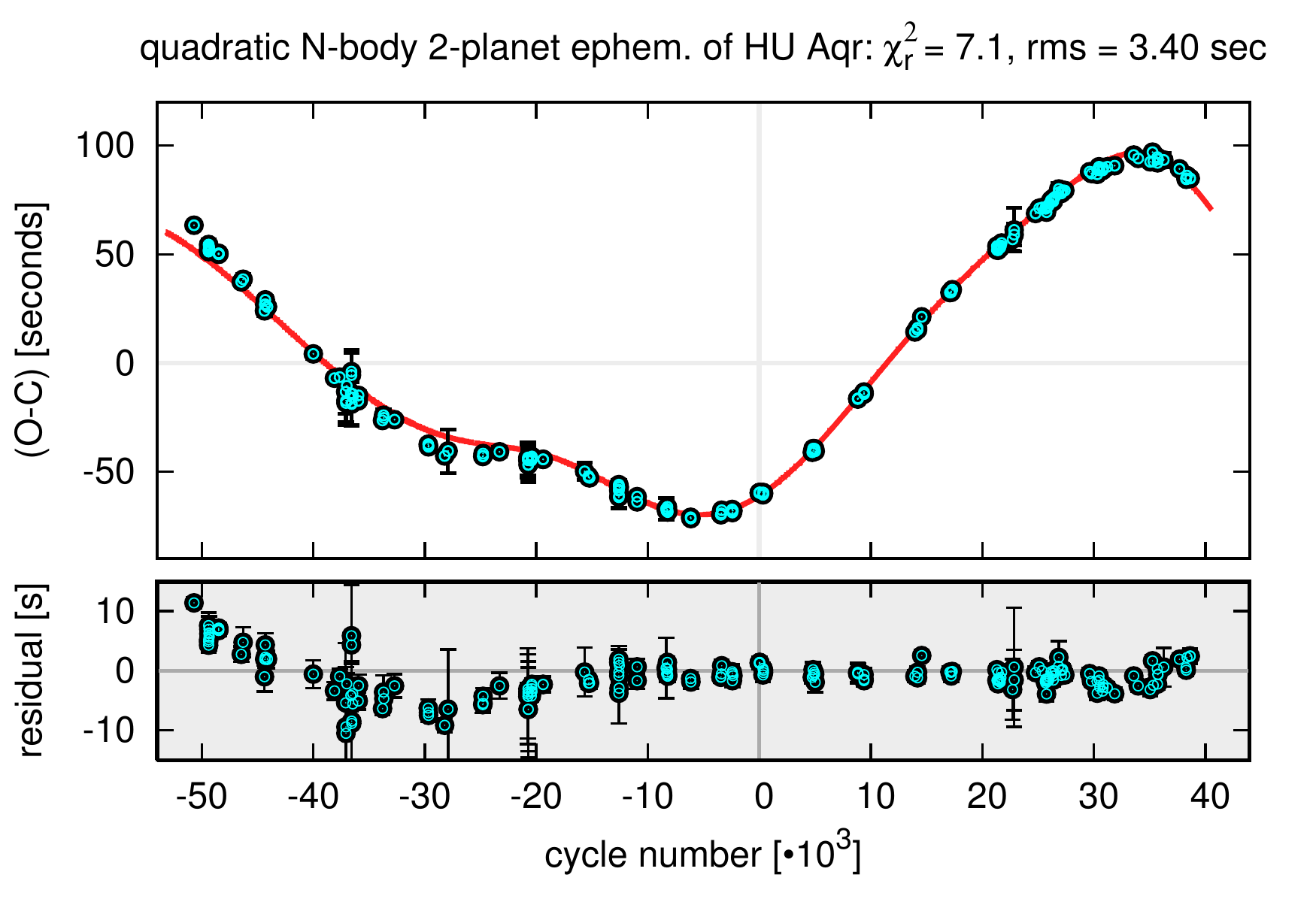}}
}
\caption{
Best-fitting 2-planet Jacobian solution (Fit JQ), {\em left panel}. This model
include the quadratic term, see Table~\ref{tab:tab3} for parameters of this
fit. {Shaded curves are for the individual LTT components, respectively}. {\em
Right panel}: $N$--body synthetic curve for the osculating initial condition
derived through the formal transformation of the Jacobian elements JQ to the
$N$-body Cartesian osculating frame, centred at the CM of the binary.
}
\label{fig:fig4}
\end{figure*}

%
The resulting set of HA-derived solutions to the 2-planet quadratic ephemeris
is illustrated as projections of their parameters onto selected planes in
Fig.~\ref{fig:fig3}. This set reveals a~shallow minimum of $\Chi \sim 1.0$ and
an rms$\sim 1.8$~s. However, the set of formal $1\sigma$ solutions providing
$\Chi<1.04$ forms a narrow ``valley'' in particular planes. This indicates
that the Keplerian model is unconstrained. The $1\sigma$--solutions might be
divided into two groups. The first group is characterised with $K_{\idm{b,c}}
\sim 30$--$60$~s and bounded orbital periods locating putative 2-planet
systems close to the $3:2$~MMR. The best-fitting model JQ of this type is
illustrated in the right-hand panel of Fig.~\ref{fig:fig4}, see
Tab.~\ref{tab:tab3}. The second group of models reveals an apparently well
bounded inner orbital period $P_{\idm{b}} \sim 4800$~days, but the outermost
orbital period $P_{\idm{c}}$ is unconstrained, and it may be as large as
36,000~days and longer. This is correlated to the LTT semi-amplitude
$K_{\idm{c}}$ up to $600$~s. The inferred mass of the outermost companion may
be as large as $80\,\mjup$ and larger. Formal uncertainties of Fit~JQ in
Table~\ref{tab:tab3} are determined with the help of the MCMC analysis of the
optimisation model including the $\sigma_f$ correction as a free parameter, in
accord with Eq.~\ref{eq:Lfun}.

\begin{figure*}
\centerline{
 \hbox{\includegraphics[ width=0.82\textwidth]{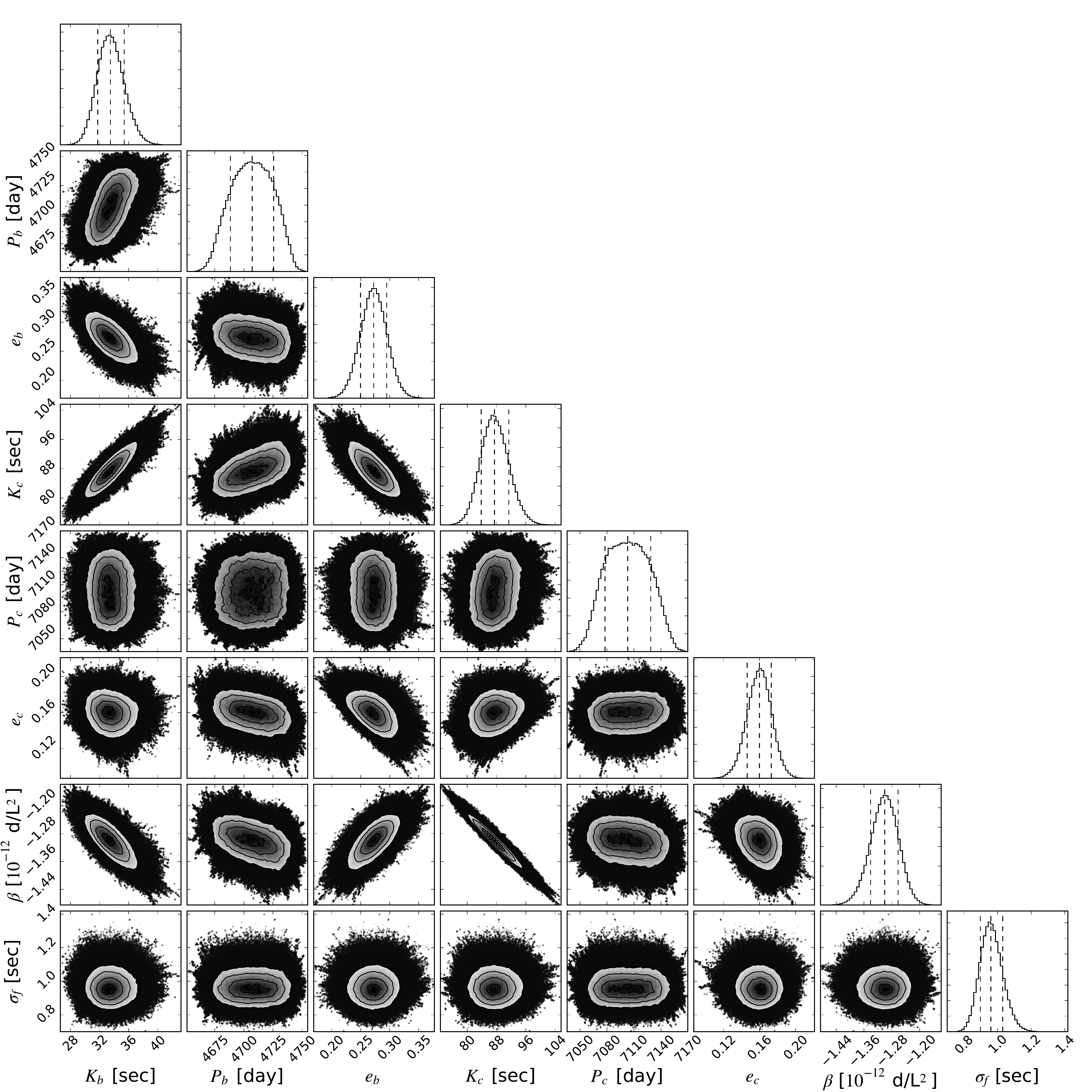}}
}
\caption{
One-- and two--dimensional projections of the posterior probability
distribution for the best-fitting kinematic model JQ with the parabolic
ephemeris, for a~few selected parameters (Table~\ref{tab:tab3}). 
}
\label{fig:fig5}
\end{figure*}

An inspection of Fig.~\ref{fig:fig3} reveals strong correlations between the
parameters. This is particularly well visible in the $(P_{\idm{c}},
e_{\idm{c}})$-- and $(P_{\idm{c}},\beta)$--planes. A similar correlation is
also found by \cite{Marsh2014} for the 2-planet parabolic ephemeris of NN~Ser
{(let us note that the residuals to the linear ephemeris of NN Ser are
qualitatively similar to the HU Aqr case, recently showing a phase of fast
increase after initially quasi-periodic oscillations)}. We investigated its
nature with an independent method, by performing the MCMC analysis of Fit JQ
(Fig.~\ref{fig:fig5}). Besides the $(P_{\idm{c}}, e_{\idm{c}})$-- and
$(P_{\idm{c}},\beta)$--correlations, there is also a~strong correlation
between the orbital periods and the time of pericenter passage for each orbit
(not shown here). Moreover, even in the smallest range, the $\beta$ term has
a~large magnitude of $-1 \times 10^{-12}$~day$\cdot L^{-2}$, making the
kinematic model questionable due to the unknown origin \corr{of such a} large
period derivative. 

\begin{table}
\centering
\caption{
Keplerian parameters in accord with 2-planet LTT fit model with parabolic
ephemeris to all available data in Table~\ref{tab:tab5}. Synthetic curves of
these solutions with mid--egress times are illustrated in Fig.~\ref{fig:fig4}.
Numbers in parentheses are for the uncertainty at the last significant digit.
Total mass of the binary is 0.98~$M_{\sun}$~\citep{Schwope2011}.
$T_0=$JD~2,453,504.8882940 is the adopted osculating epoch coinciding with the
$L=0$ epoch, close the middle of observational window.
}
\begin{tabular}{ccc}
 \hline
\label{tab:tab3}
Parameters & Fit JQ &  Fit JL \\
\hline  
$K_\idm{b}$~[s]       &  33.6 $\pm$ 1.9        & 2765.6 $\pm$ 331.5 \\
$P_\idm{b}$~[day]     & 4707.4 $\pm$ 19.0      & 8029.4 $\pm$ 18.2 \\
$e_\idm{b}$           &  0.270  $\pm$ 0.022    & 0.317 $\pm$ 0.015  \\
$\omega_\idm{b}$~[deg] 
                      &  44.640  $\pm$ 8.0     & 206.2 $\pm$ 4.1 \\
$T_\idm{b}$~[$T_0$+]  &  3845.4 $\pm$ 357.35   & 2917.4 $\pm$ 5524.6 \\
$a_\idm{b}$~[au]      &   5.48                 & --- \\
$m_\idm{b}\sin i$~[$\mjup$]
                      &    12.7                & --- \\
\hline
$K_\idm{c}$~[s]       &    87.7 $\pm$ 4.0      & 2860.7 $\pm$  337.2 \\
$P_\idm{c}$~[day]     &  7101.7 $\pm$ 25.6     & 8298.8 $\pm$ 19.8 \\
$e_\idm{c}$           & 0.159 $\pm$ 0.014      & 0.336 $\pm$ 0.014 \\
$\omega_\idm{c}$~[deg]  
                      & 346.2 $\pm$ 5.2        & 25.03 $\pm$ 4.0 \\
$T_\idm{c}$~[$T_0$+]  & 3784.5 $\pm$ 1576.0    & 3714.1 $\pm$ 5766.5 \\
$a_\idm{c}$~[au]      &     7.14               & --- \\
$m_\idm{c}\sin i$~[$\mjup$]&    25.8           & --- \\
\hline
$P_\idm{bin}$~[day]   & 0.8682036931(4)        &  0.0868203618(6) \\
$t_0$ [BJD 2,453,504.0+]  & 0.88899(1)         &  0.88742(15) \\ 
$\beta$ [$\times10^{-12}\,$day$\cdot L^{-2}$]  & -1.30(4)   & --- \\
$\sigma_f$ [s]        &  0.92(8)               &   1.48(9)\\
\hline
$N_\idm{obs}$ data    & 205                    &  205 \\
$\Chi$                & 0.96                   &  1.04 \\ 
rms [s]               & 1.72                   & 2.02  \\
\hline
\end{tabular}
\end{table}
\subsection{Linear ephemeris, kinematic 2-planet model}
\begin{figure}
\centerline{\hbox{\includegraphics[ width=0.48\textwidth]{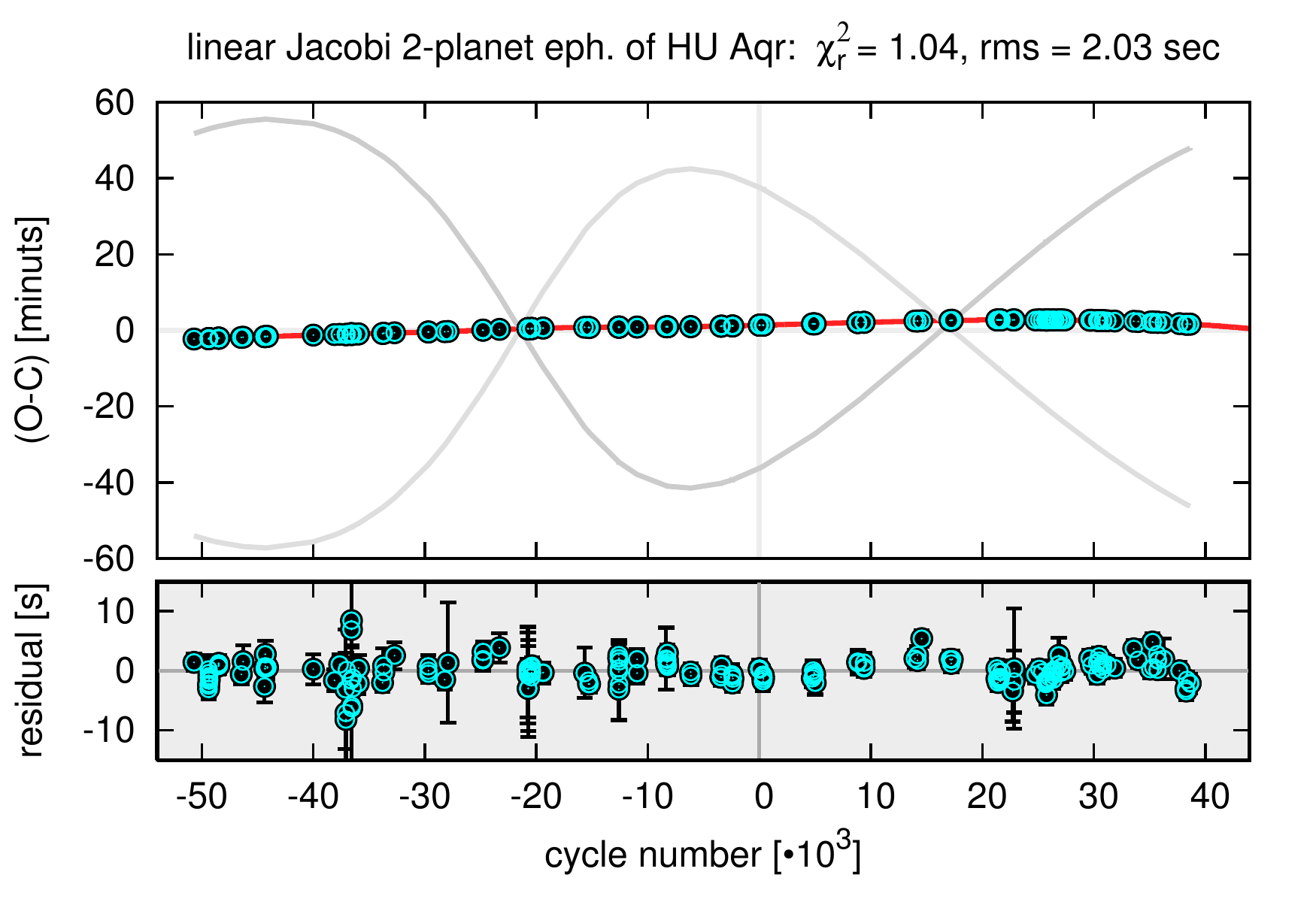}}}
\caption{
The synthetic curve for the best-fitting 2-planet model \corr{JL} with the linear
ephemeris (see Tab.~\ref{tab:tab3} for its parameters). Shaded curves are for
individual signals of both companions, respectively.
}
\label{fig:fig6}
\end{figure}

For reference, we also investigated the 2-planet linear ephemeris model,
{which is a very ``attractive'' variant of the LTT hypothesis. A stable system
consistent with the linear ephemeris and observations would essentially solve
the problem of large orbital period derivative $\beta$.} Unfortunately, the
linear model is even less constrained than the quadratic ephemeris case. The
best-fitting solution found after an extensive HA search {provides $\Chi\sim
1.04$ and an rms $\sim 2.1$~s}. It reveals similar orbital periods
$P_{\idm{b,c}} \sim8,300$~days. Simultaneously, the semi-amplitudes
$K_{\idm{b,c}}$ may be as large as $\sim 1$~{\em hour}. Such huge
semi-amplitudes imply companion masses in the range characteristic for red
dwarfs and massive stars. This is of course non-physical outcome of the
mathematical model. The best-fitting solution JL is illustrated in
Fig.~\ref{fig:fig6}, see Tab.~\ref{tab:tab3} for its parameters. A particular
orientation of orbits with very similar periods, and anti-aligned planets
provides the same (O-C) in wide ranges of masses. The same kind of degeneracy
of kinematic 2-planet model with the linear ephemeris is present in the older
data, and we discussed this problem in \citet{Gozdziewski2012}. Looking at the
residuals (bottom panel in Fig.~\ref{fig:fig6}) we notice systematic, though
apparently small deviation from zero at the end of the observing window. All
best-fitting models to the linear ephemeris exhibit measurements similarly
outlying from their synthetic solutions. 

\begin{figure*}
\centerline{\hbox{\includegraphics[ width=0.82\textwidth]{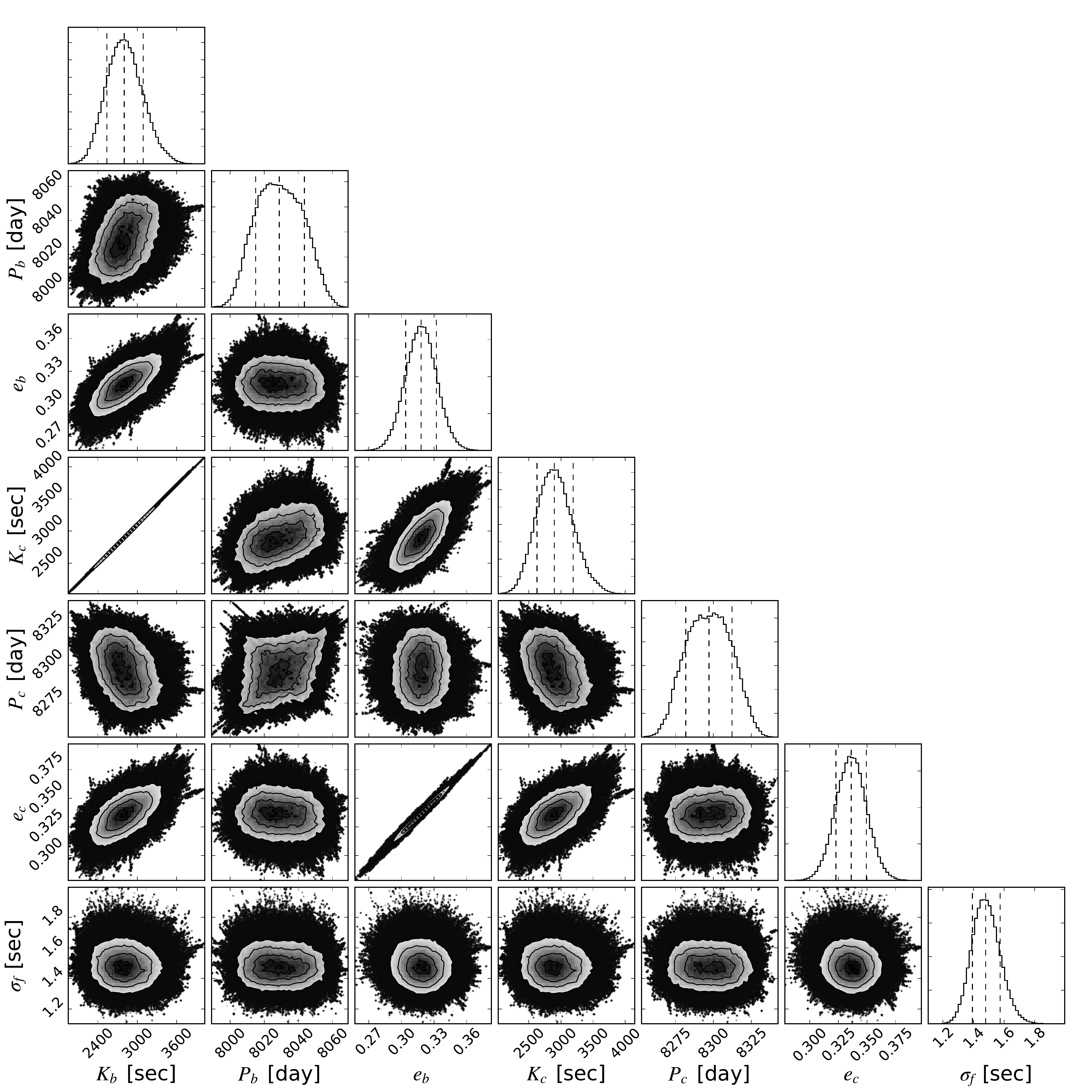}}}
\caption{
Projections of the MCMC posterior probabilities for selected parameters of the
best-fitting solution JL (see Table~\ref{tab:tab3} for elements of this fit).
Note extremely large semi-amplitudes of individual LTT signals, strong
correlations between similar orbital periods $P_{\idm{b,c}}$ and between
semi-amplitudes $K_{\idm{b,c}}$.
}
\label{fig:fig7}
\end{figure*}

We found that the JL parameter correlations are much stronger than for the
quadratic ephemeris. This is illustrated by the projections of the MCMC
posterior probability distributions for a~few selected model parameters in
Fig.~\ref{fig:fig7}. A reliable determination of their uncertainties is very
difficult unless the model is not re-parametrised in some particular way. The
JL solution is quoted solely to show extreme values of its non-physical
parameters and its degenerate character.

%
\subsection{Conversion of Keplerian elements to the $N$-body frame}
%
Following a common approach in the recent literature, we should check whether
the best-fitting planetary systems are dynamically stable. Moreover, recalling
close orbital periods in the best-fitting solutions JQ and JL, kinematic
models may be invalid at all, due to significant mutual gravitational
interactions between the planets. This will be shown in the next section. 

\begin{figure}
\centerline{
\hbox{\includegraphics[ width=0.48\textwidth]{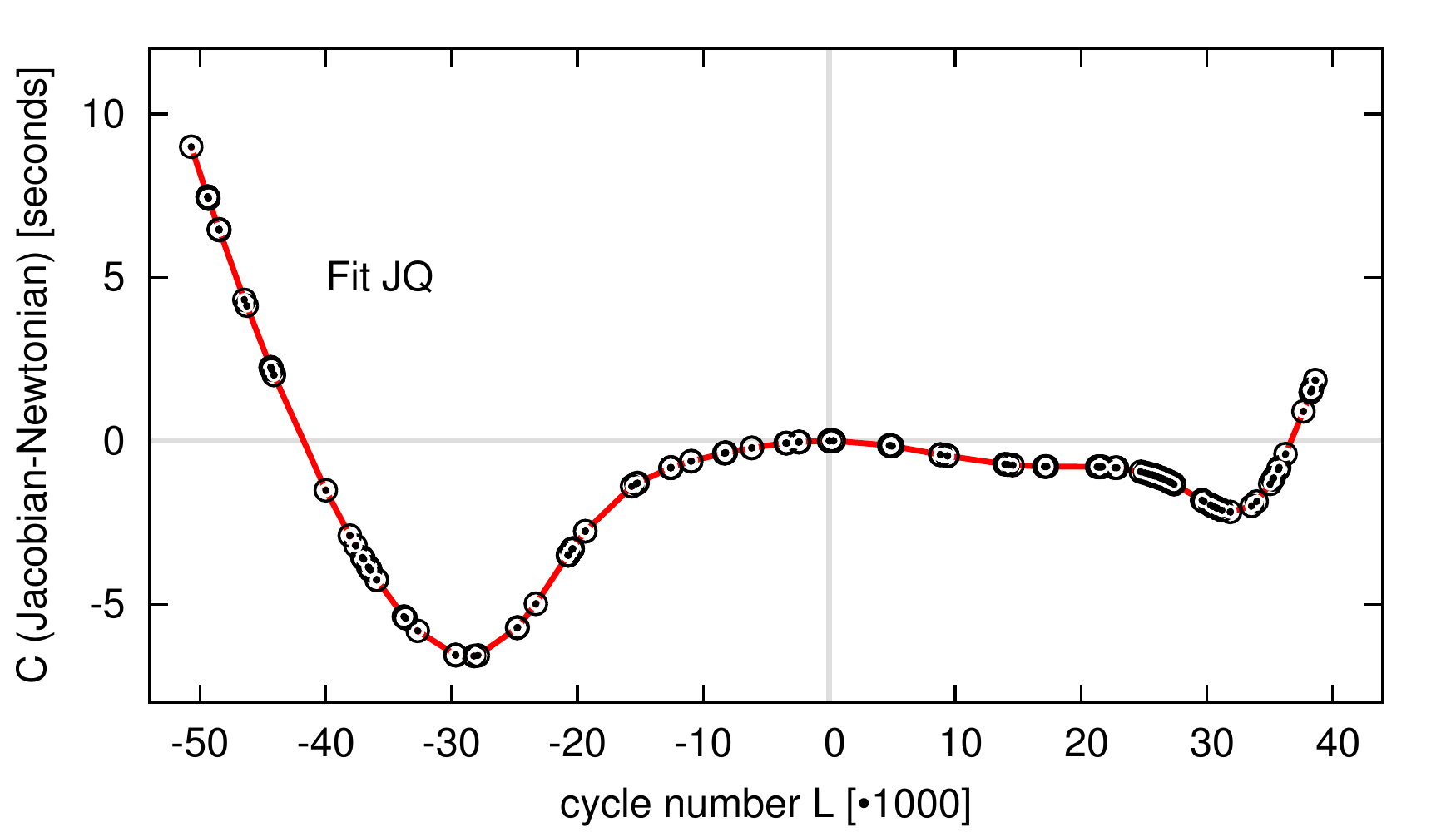}}
}
\caption{
Differences \corr{of the LTT signals derived}
from Jacobian, kinematic solutions JQ
(Tab.~\ref{tab:tab3}) and from respective, osculating $N$-body model, integrated
numerically with the inferred initial condition at the osculating epoch of the
$L=0$ cycle. The $L=0$ cycle is centred roughly at the middle of the
observational window.
}
\label{fig:fig8}
\end{figure}

To accomplish numerical $N$-body integrations, we need to transform the
initial conditions from the Jacobian, kinematic frame, to the $N$-body
Cartesian coordinates at the osculating epoch $T_0${\footnote{Note, that in
the literature, kinematic elements derived as ``raw'' parameters of the (O-C)
model, Eq.~\ref{eq:epla}, are usually interpreted incorrectly as 
osculating, \corr{{\em Keplerian, astrocentric}} elements w.r.t. the CM of the binary.}}. {An example
of such a conversion for Fit JQ is illustrated in Fig.~\ref{fig:fig4}. The
right panel shows systematic trends of the residuals to the $N$-body solution
outside the $T_0$ epoch. Indeed, a direct comparison of the synthetic signals
derived from the kinematic and $N$-body models differ by $\pm 10$~seconds
(Fig.~\ref{fig:fig8}). This might be still considered as a~subtle difference,
however, the $N$-body osculating initial condition derived from Fit JQ
provides $\Chi>7.29$, though $\Chi\sim 1$ for the original, source kinematic
model}. This means that the transformed kinematic initial conditions are
poorly consistent with observations. A similar discrepancy of Newtonian and
kinematic 2-planet models has been shown in \citet{Marsh2014} for the (O-C) of
NN~Serpentis. These differences for HU~Aqr are 10~times larger, {likely due to
larger planetary masses and smaller separation of orbits}. In some parts of
this window they may be compared to the signal itself. Recalling a possibility
of massive companions, this shows that the LTT signal of HU~Aqr cannot be
properly modelled even in terms of the Jacobian, refined kinematic
formulation.

Fortunately, the gathered huge sets of $\sim10^6$ Jacobian fits with $\Chi~<~9$ may be
still used as relatively accurate approximations of the proper osculating
elements. These solutions were further refined in terms of the exact $N$-body
model. Such an approach is helpful for CPU--efficient and quasi-global
exploration of the parameter space of the Newtonian models.
%
\subsection{The $N$-body 2-planet model with parabolic term}
\label{subsec:nq2model}
\begin{figure*}
\centerline{
 \hbox{\includegraphics[ width=0.44\textwidth]{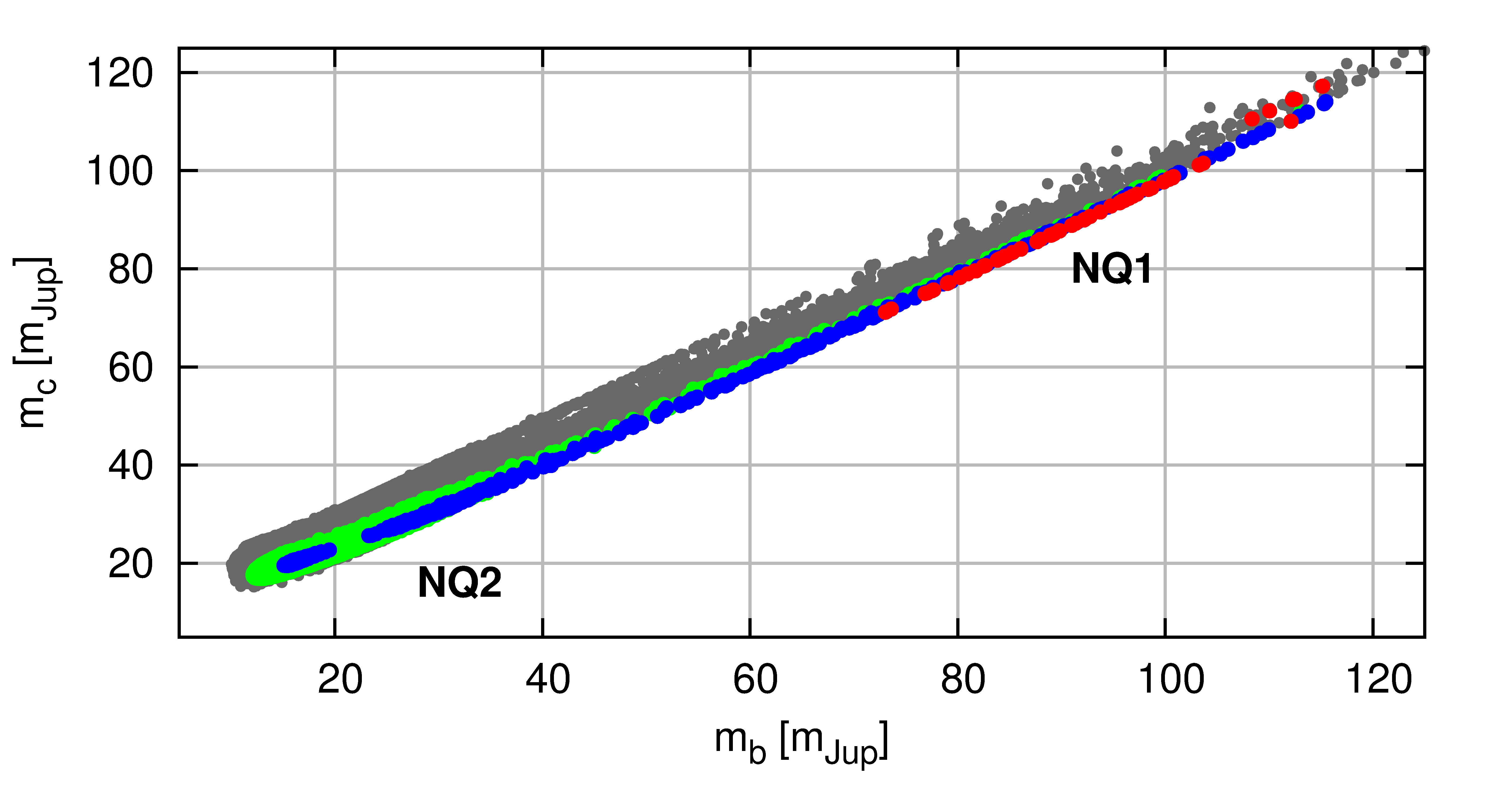}}
 \hbox{\includegraphics[ width=0.44\textwidth]{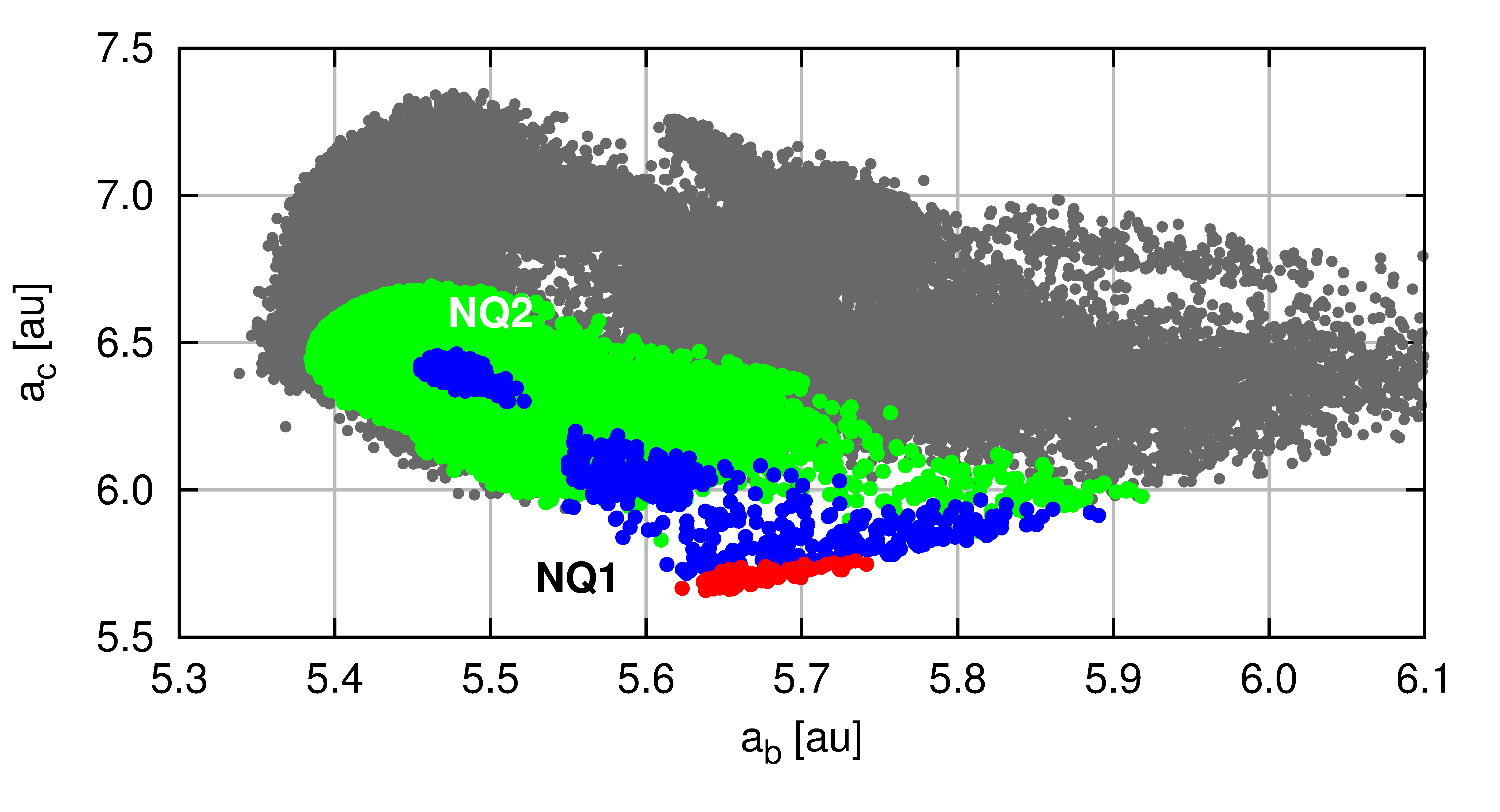}}
}
\centerline{
 \hbox{\includegraphics[ width=0.44\textwidth]{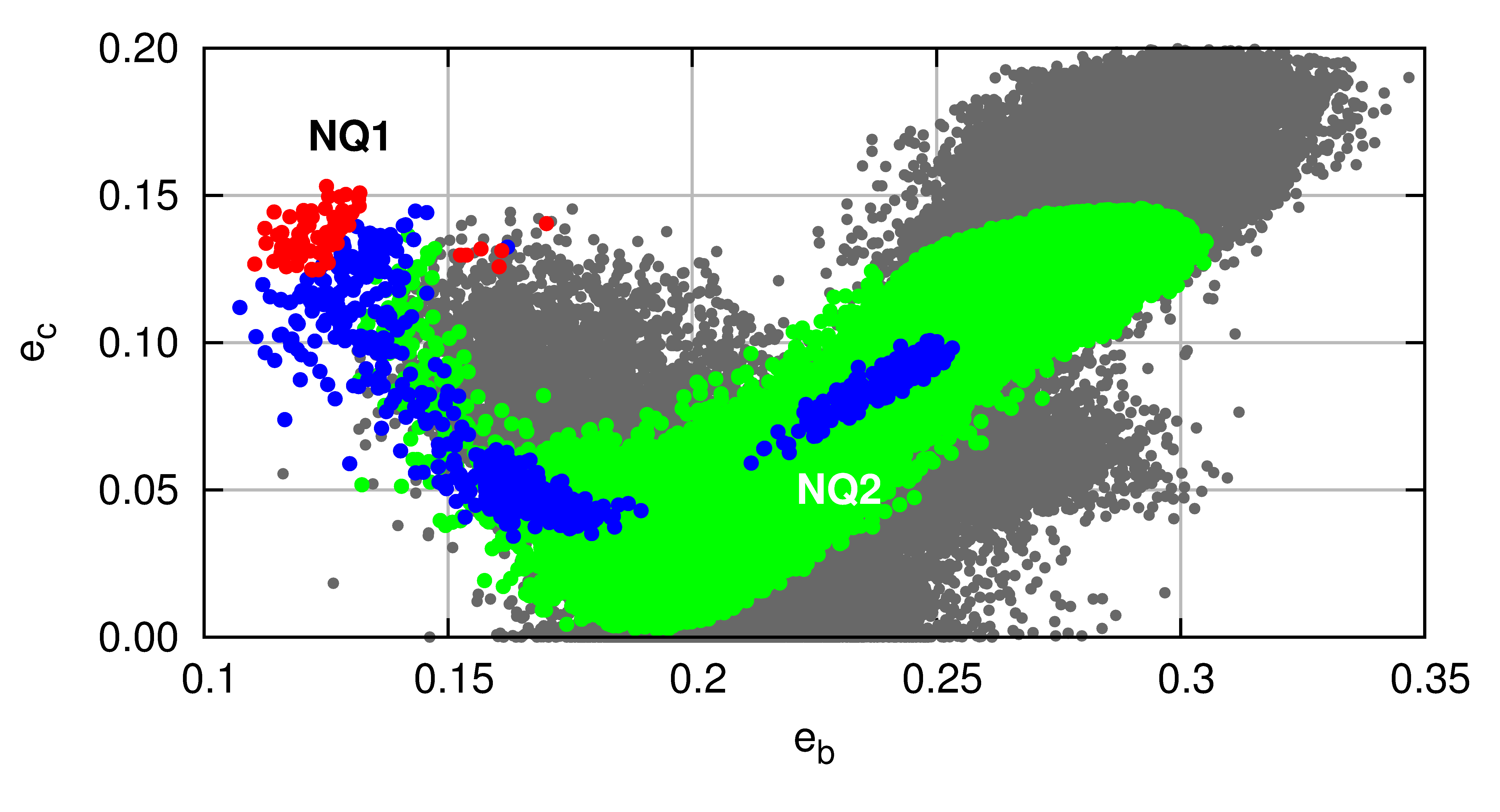}}
 \hbox{\includegraphics[ width=0.44\textwidth]{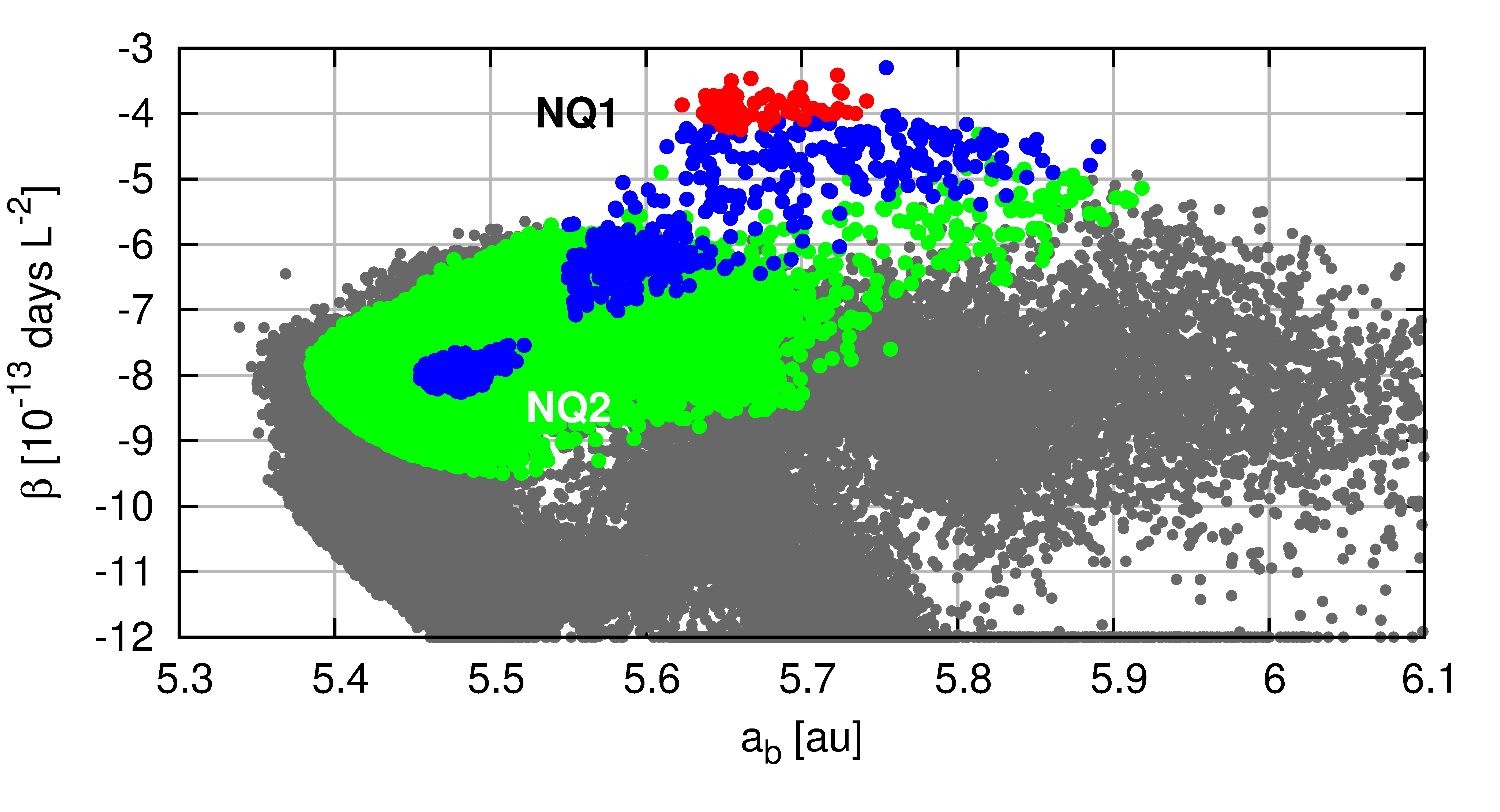}}
}
\centerline{
 \hbox{\includegraphics[ width=0.44\textwidth]{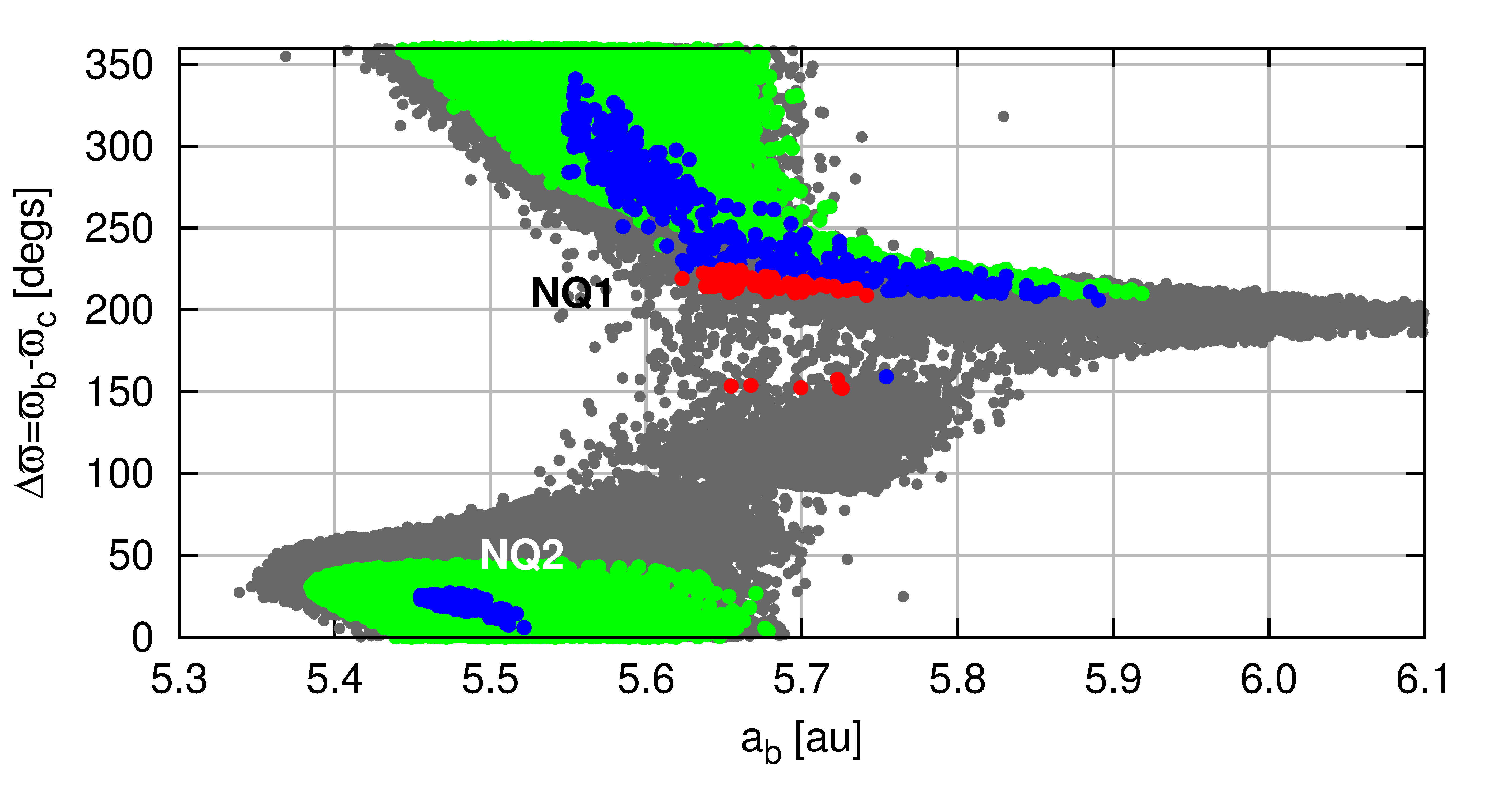}}
 \hbox{\includegraphics[ width=0.44\textwidth]{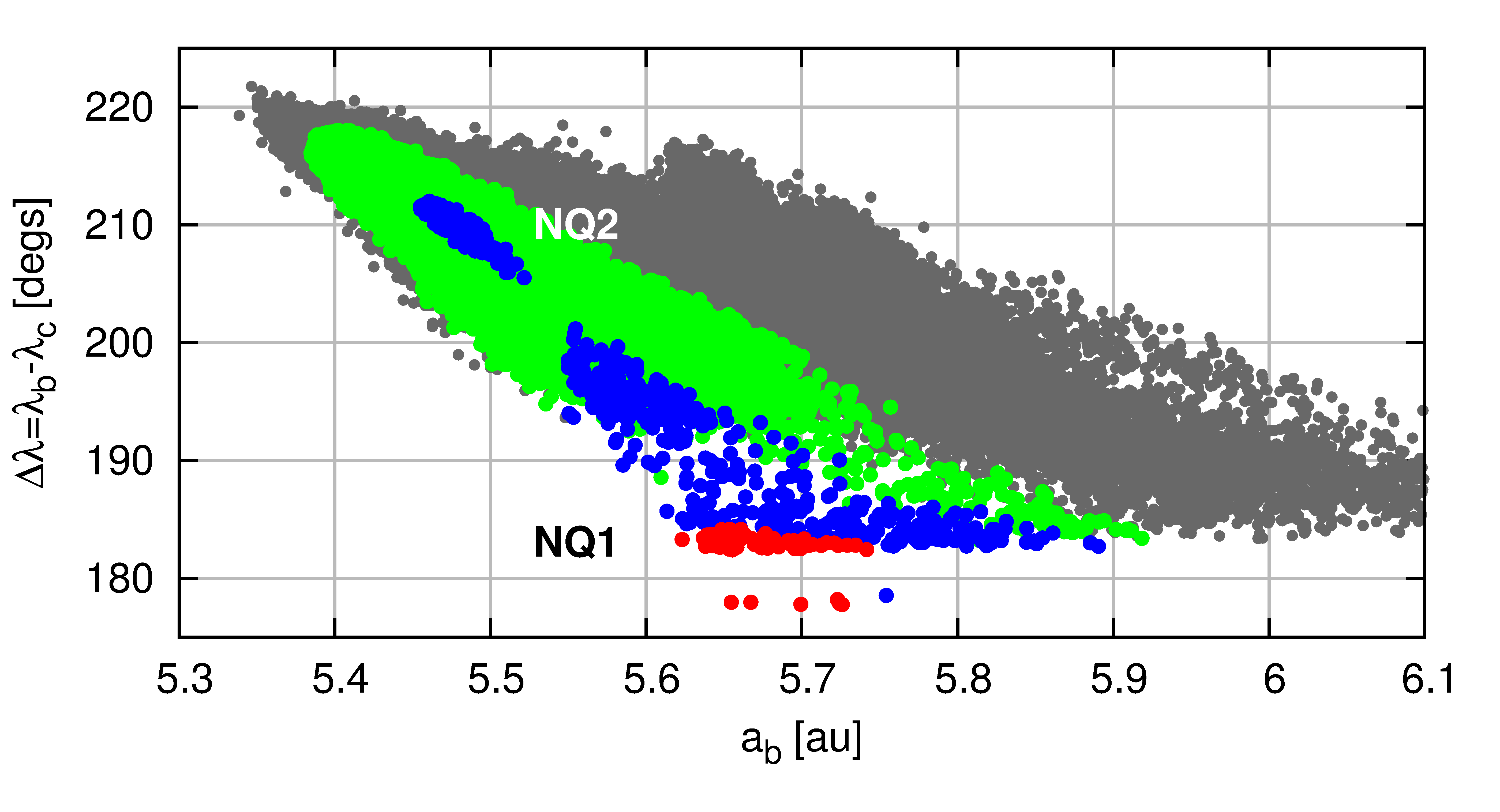}}
}
\caption{
Best-fitting 2-planet Newtonian solutions with the quadratic term projected
onto planes of orbital elements at the osculating epoch $T_0$. Red filled
circles are for the best-fitting models with $\Chi<0.96$, blue filled circles
are for $\Chi<1.02$, green filled circles are for $\Chi<1.06$ (roughly
$3\sigma$ of Fit NQ1), and grey filled circles are for $\Chi<1.17$,
respectively. Solutions NQ1 and NQ2 displayed in Table~\ref{fig:fig4} are
labelled.
}
\label{fig:fig9}
\end{figure*}

\begin{figure*}
\centerline{
 \hbox{\includegraphics[ width=0.48\textwidth]{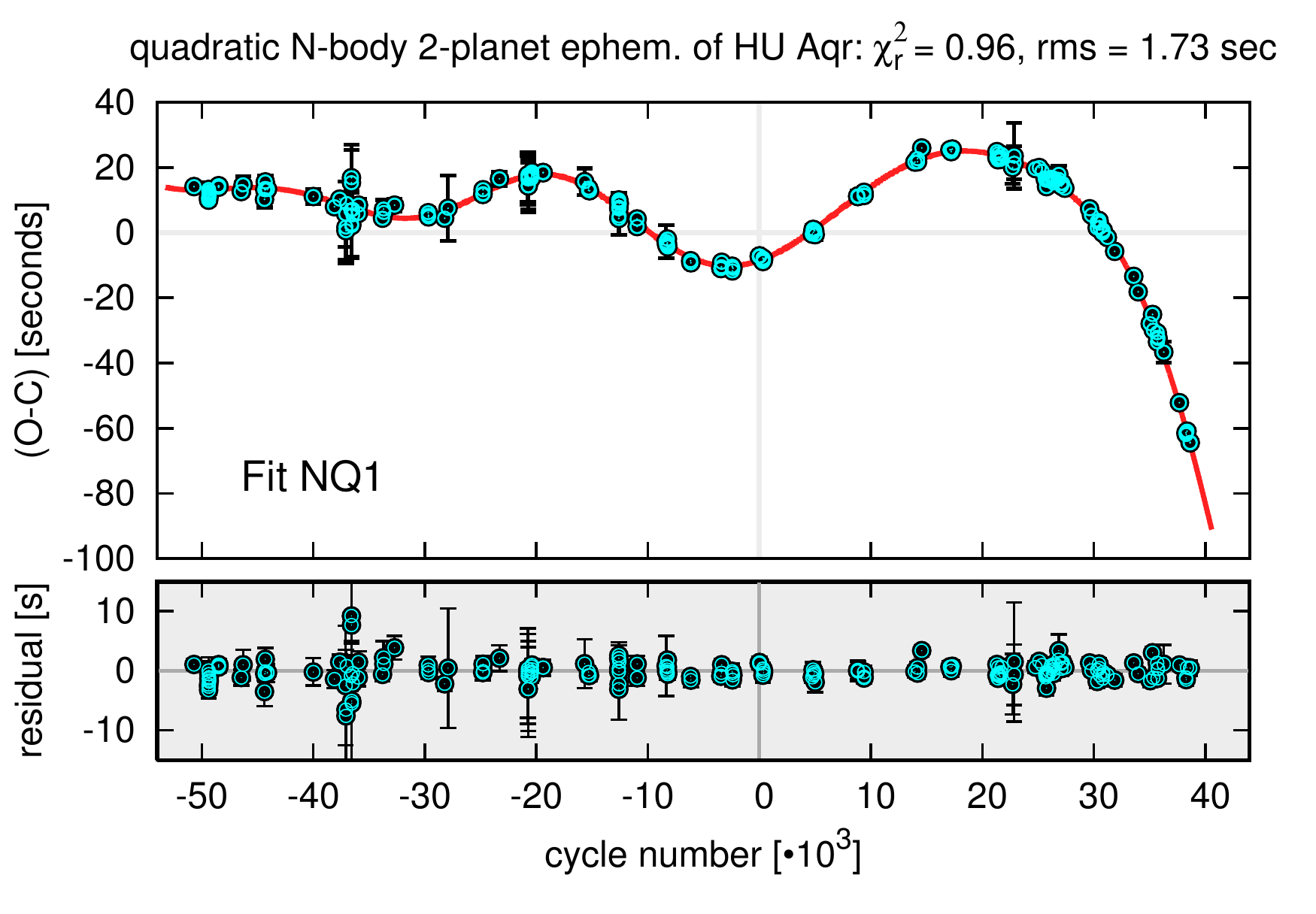}}
 \hbox{\includegraphics[ width=0.48\textwidth]{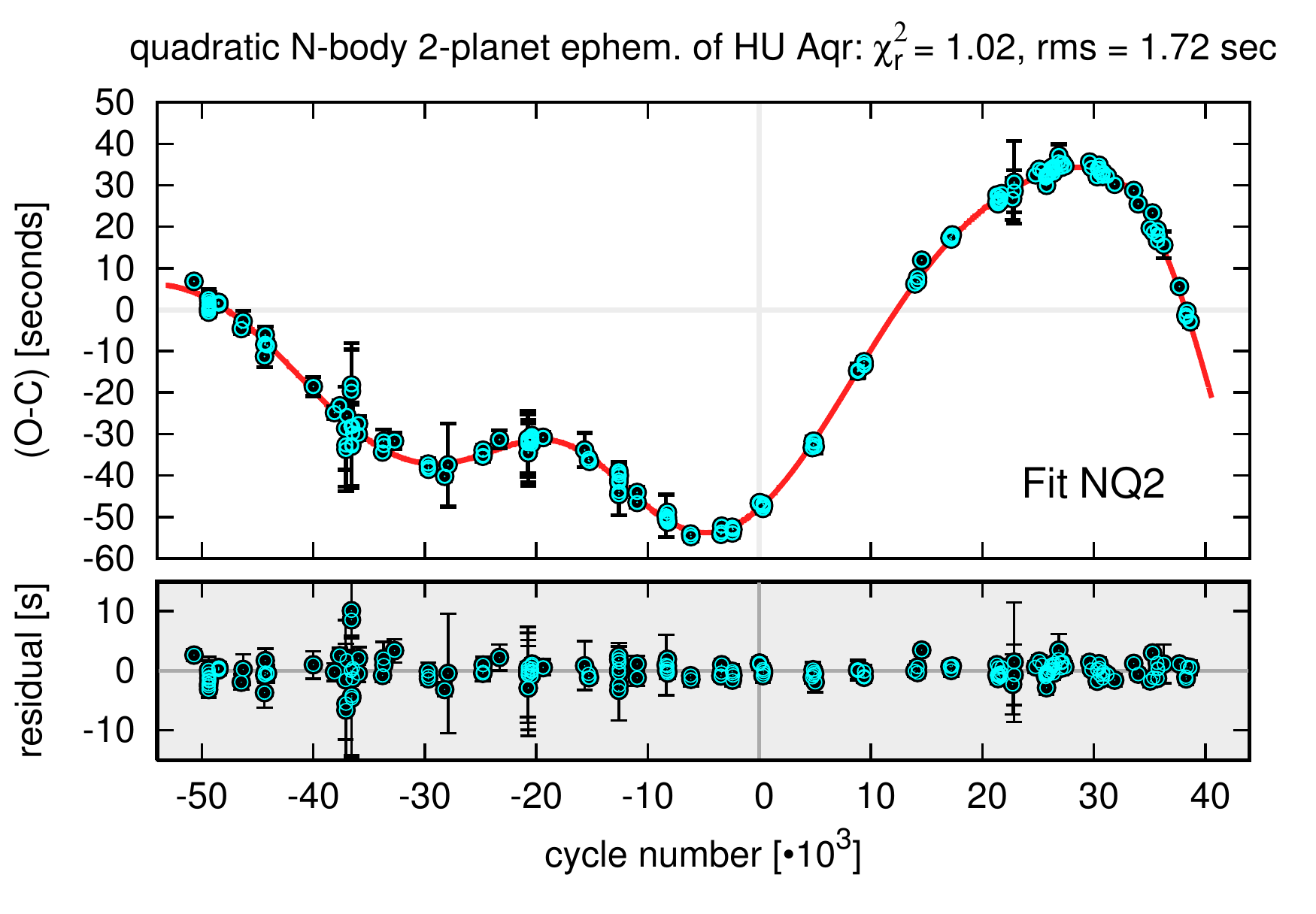}}
}
\caption{
Best-fitting Newtonian 2-planet quadratic ephemeris models: Fit NQ1, {\em left
panel}, and Fit NQ2, {\em right panel}, respectively. See Table~\ref{tab:tab4}
for parameters of these solutions.
}
\label{fig:fig10}
\end{figure*}

The refined set of $N$-body models with parabolic ephemeris is illustrated in
Fig.~\ref{fig:fig9}. Green filled circles in this figure encode solutions
$\Chi<1.06$ (roughly $3\sigma$ of the best-fitting model NQ1 displayed in
Table~\ref{tab:tab4}). This distribution of best-fitting solutions is
bi-modal, with two types of planetary configurations. The first type is close
to the best-fitting model NQ1 providing $\Chi\sim 0.96$ and characterised by
very similar semi-major axes of both companions, $a_{\idm{b,c}} \sim 5.7$~au,
and with strongly correlated and unconstrained masses, drawn up to
$120~\mjup$. Similar semi-major axes indicate the 1:1~MMR. The second, shallow
minimum of $\Chi \sim 1.02$ is for solutions around the 3:2~MMR (Fit NQ2),
since the semi-major axes are close to $\sim 5.5$~au, and $\sim 6.5$~au,
respectively. These best-fitting models are illustrated in
Fig.~\ref{fig:fig10}.

The MCMC analysis reveals that though a clear correlation persists between the
outermost period and the period derivative term, this correlation is
significantly weaker than in the kinematic model. This parameter has a large
magnitude of $\beta \simeq -4 \times 10^{-12}$~day\,$L^{-2}$ and $\beta \simeq
-7 \times 10^{-12}$~day\,$L^{-2}$, respectively. The MCMC derived posterior
probability distribution histograms (not show here) confirmed in independent
way that both solutions, derived and illustrated through the HA projections in
Fig.~\ref{fig:fig9}, are relatively well bounded in the parameter space,
except for a very strong correlation of masses in Fit~NQ1, similar to
correlation of semi-amplitudes in Fit~JL.

\begin{table}
\centering
\caption{
Newtonian parameters of two best-fitting 2-planet LTT models with parabolic
ephemeris, NQ1 and NQ2, respectively. Synthetic curves with mid--egress times
are illustrated in Fig.~\ref{fig:fig10}. Digits in parentheses are for the
uncertainty at the last significant place. Total mass of the binary is
0.98~$M_{\sun}$~\citep{Schwope2011}.
}
\begin{tabular}{cccc}
 \hline
\label{tab:tab4}
Parameters & Fit NQ1 & Fit NQ2  \\
\hline
$m_{\idm{b}}$~[$\mjup$]  & 96.2  $\pm$ 9.4   & 16.8  $\pm$  1.0   \\
$a_{\idm{b}}$~[au]       & 5.68  $\pm$ 0.06  & 5.48  $\pm$ 0.06  \\
$e_{\idm{b}}$            & 0.147 $\pm$ 0.015 & 0.230 $\pm$ 0.028 \\
$\omega_{\idm{b}}$~[deg] & 250.8 $\pm$ 7.3   & 92.3  $\pm$ 7.5  \\
$M_{\idm{b}}$~[deg]      & 170.3 $\pm$ 7.7   & 115.6 $\pm$ 4.1  \\
\hline
$m_{\idm{c}}$~[$\mjup$]
                         & 98.3  $\pm$ 9.5   & 20.8 $\pm$ 1.3   \\
$a_{\idm{c}}$~[au]       & 5.67  $\pm$ 0.06  & 6.38 $\pm$ 1.56   \\
$e_{\idm{c}}$            & 0.123 $\pm$ 0.011 & 0.083 $\pm$ 0.027  \\
$\omega_{\idm{c}}$~[deg] & 103.0 $\pm$ 8.1 & 72.6 $\pm$ 9.2  \\
$M_{\idm{c}}$~[deg]      & 140.8 $\pm$ 9.0 & 286.5  $\pm$ 5.63   \\
\hline
$P_{\idm{bin}}$~[day]  & 0.0868203933(9)  & 0.0868203796(1)  \\
$\Delta t_0$~[BJD 2,453,504.0+]  & 0.88838(3)  & 0.88884(1)  \\
$\beta$ [$\times 10^{-13}$ day\,$L^{-2}$]   & -3.7(2)   &-7.9(5) \\
$\sigma_f$~[s]           &    0.94(7) &  0.94(8)   \\
\hline
$N_{\idm{obs}}$ data points & 205     & 205   \\
$N_{\idm{par}}$ free parameters       & 14  & 14 \\
$\Chi$                   & 0.96       &  0.96  \\ 
rms [s]                  & 1.73        & 1.78  \\
\hline
\end{tabular}
\end{table}

We also computed the Newtonian models with the linear ephemeris. In this case,
only one minimum of $\Chi$ is apparent in the region of the 1:1~MMR, with both
semi-major axes around $\sim 7$~au. However, all orbital parameters are spread
over wide ranges. There is also a strong correlation between unconstrained
companion masses, which reach the non-physical stellar mass range, similar to
the kinematic 2-planet linear ephemeris. This correlation has the same
geometric source as in the Keplerian model, since it appears due to
particularly anti-aligned orbits with similar semi-major axes. Therefore, the
best-fitting, coplanar 2-planet models with the linear ephemeris must be also
considered as highly unconstrained and non-physical. 

%
%
\subsection{Stability analysis of $2$-planet solutions}
%
\label{sec:stability}
At a stage of the $N$-body correction of Keplerian models, we verified the
dynamical stability of all solutions with final $\Chi<4.0$, by the direct
numerical integration of the $N$-body equations of motion with the
\corr{Bulirsch-Stoer} algorithm. To quantify the dynamical stability, we determined
the crossing time of orbits (the Event Time, $T_\idm{E}$ from hereafter) for
\corr{a time span} of $\sim 10^6$ outermost orbital periods. If $T_\idm{E}$ is
shorter than this limiting integration time, then this indicates a close
encounter between components or the ejection of a planet from the system. We
did not find any stable solutions {in the set of $\sim 10^6$ models} with the
parabolic ephemeris, close to the best-fitting solutions NQ1 and NQ2 (as
illustrated in Fig~\ref{fig:fig9}), except for just a~few hierarchical
configurations with $\Chi>1.6$. Such stable solutions are characterised by the
outermost semi-major axis $a_{\idm{c}} \sim 15$~au and a large mass of the
outer companion $\sim 69\,\mjup$, i.e., in a region of the parameter space
separated by a few $\sigma$ from the best-fitting models.

This still does not prove that stable configurations with reasonably small
$\Chi$ do not exist in some regions of the parameter space, consistent with
observations. {For instance, the angular elements of such solutions might be
systematically displaced from a stable MMR region
\citep[e.g.][]{Gozdziewski2001,Gozdziewski2014}}. Therefore, to understand the
observed instability, we carried out extensive Monte Carlo experiments. For
the 2-planet quadratic ephemeris solutions, we selected grids of
$400\times400$ and $400\times800$ points in the semi-major axes plane
$(a_{\idm{b}},a_{\idm{c}})$, spanning $a_{\idm{b}} \in [5,7]$~au, $a_{\idm{c}}
\in [5,8]$~au for fit NQ1, and $a_{\idm{b}} \in [4.5,6.5]$~au, $a_{\idm{c}}
\in [4.5,9.5]$~au for fit NQ2, respectively. Next, at each point of the given
grid, we computed the Mean Exponential Growth factor of Nearby Orbits (MEGNO
from hereafter) for up to $1$~Myr (i.e., up to a few $10^4$ outermost orbital
periods). MEGNO \citep{Cincotta2003,Gozdziewski2001} is a numerical algorithm
making it possible to estimate efficiently the Maximal Lyapunov Exponent and
to determine quasi-regular and chaotic solutions of the $N$-body equations of
motion. In the examined ranges of semi-major axes, 2-planet systems may be
stable only in the regime of MMR's, and chaotic solutions imply short-time
geometric instability (hence short $T_{\idm{E}}$)
\citep[e.g.][]{Gozdziewski2014}.

The MEGNO indicator was computed 2880 times at each point of a particular
$(a_{\idm{b}},a_{\idm{c}})$-grid: at each point of this grid, the argument of
pericenter $\varpi_{c} \in [0^\circ,360^\circ]$ was gradually increased from
$0^{\circ}$ to $360^{\circ}$ by $\Delta\varpi_{\idm{c}} = 0.125^{\circ}$, and
for a such a particular value of $\varpi_{\idm{c}}$, we selected random
eccentricities within $[0,e_{\idm{b,c}}+0.2)$ around a given best-fitting
solution, and random mean anomaly $M_{c} \in [0^\circ,360^\circ]$. In this
way, we obtain an extensive and exhaustive mapping of the multidimensional
parameter space. The tested sets of initial conditions for the best-fitting
models NQ1 and NQ2 contain of $\sim 4\times10^{8}$ and $\sim 9\times10^{8}$
elements, respectively. (Note that planetary masses were kept fixed in both
tests at their nominal values of NQ1 and NQ2 solutions in
Table~\ref{tab:tab4}). Then we computed MEGNO for each initial conditions and
gathered all regular (quasi-periodic) models. Such massive computations would
be hardly possible to conduct without a help of our Message Passing Interface
(MPI) code $\mu${\sc Farm} run at the {\tt cane} computer cluster (Pozna\'n
Supercomputing Centre PCSS, Poland).  

\begin{figure*}
\centerline{
     \hbox{
         \hbox{\includegraphics[ width=3.5in]{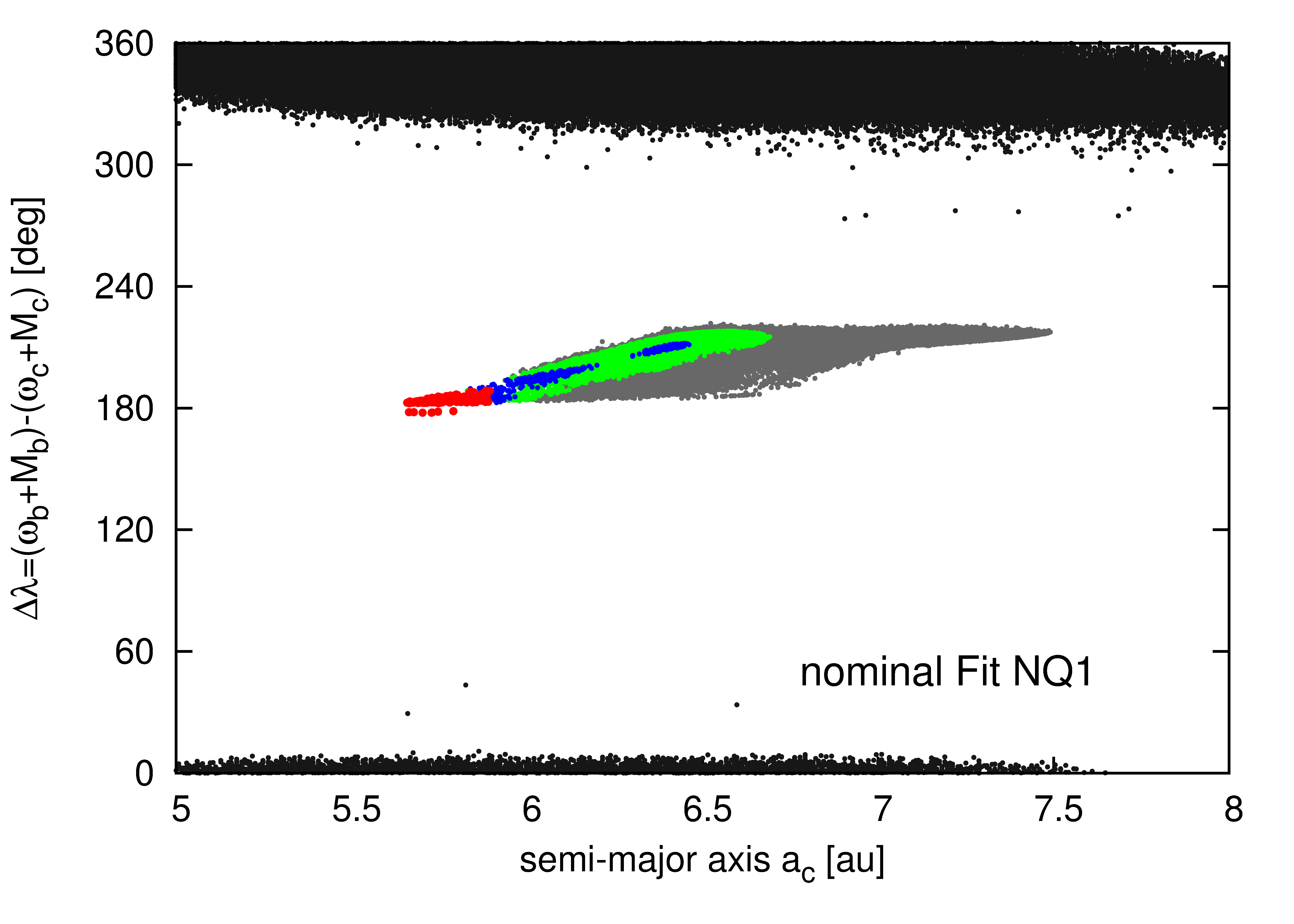}}
         \hbox{\includegraphics[ width=3.5in]{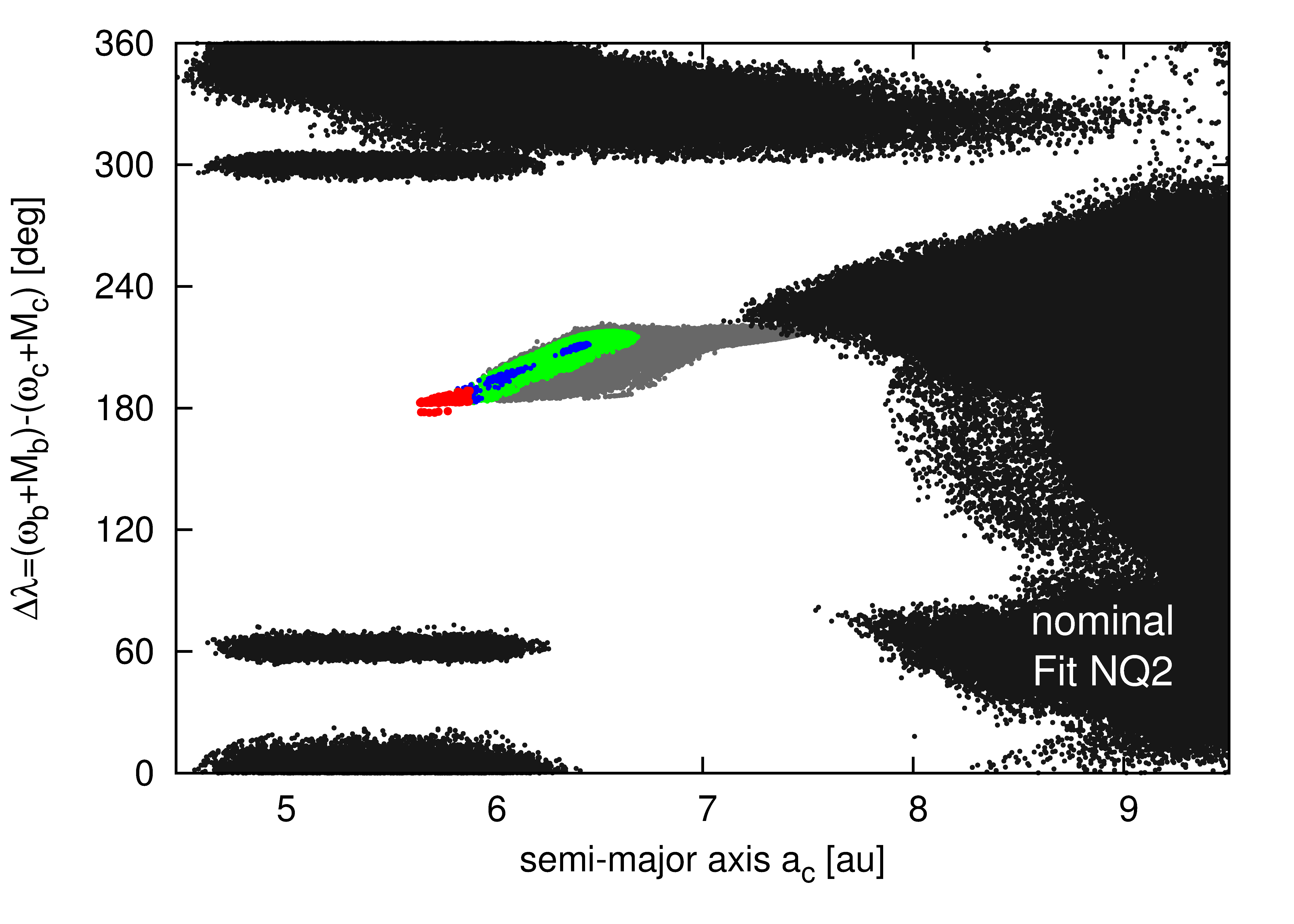}}
   }
}
\caption{
Illustration of stability analysis of the 2-planet Newtonian models NQ1 ({\em
left panel}), and NQ2 ({\em right panel}) in the
($a_{\idm{b,c}},\Delta\lambda)$--planes, where $\Delta\lambda$ is the initial
difference between the mean longitudes of the planets. Dark grey points are
for stable initial conditions, light grey and coloured points are for models
within $\Chi<1.21$ (a few $\sigma$--confidence levels of the best fitting
model NQ1). White colour is for unstable configurations. See the text for
details.
}
\label{fig:fig11}
\end{figure*}

The results are illustrated in two panels of Fig.~\ref{fig:fig11}, in the
$(a_{\idm{b,c}},\lambda_{\idm{b}}-\lambda_{\idm{c}})$-planes {around initial
orbital elements corresponding to Fits~NQ1 and NQ2, respectively}. For
reference, we overplotted the statistics of the best-fitting solutions
(Fig.~\ref{fig:fig9}). White regions in this figure mean that {\em any tested
combination} of the orbital elements, \corr{even} not necessarily consistent with the
observed (O-C), has a positive Lyapunov Exponent (therefore is unstable). The
best-fitting models are found around initial
$\Delta\lambda_{b,c}=\lambda_{\idm{b}}-\lambda_{\idm{c}} \sim 180^{\circ}$,
hence in anti-phase with possible stable models (dark, grey filled circles in
Fig.~\ref{fig:fig11}). {Both these regions are completely separated up to
$\Chi\sim 1.17$, more than a~few $\sigma$ levels}. This suggests that stable
configurations are unlikely within the parameter ranges permitted by the
observations, recalling the huge volume of initial conditions examined in both
experiments. Similar Monte Carlo tests were repeated a~few times more for
different nominal best-fitting models {obtained on the basis or earlier
observations}, and also their results are negative. We note that stable
configurations with proper $\Delta\lambda$ are possible for systems with the
outermost semi-major axis $a_{\idm{c}}$ greater than $\sim 15$~au. However,
such models are poorly consistent with the observations, implying $\Chi>1.56$,
as described above.

A similar stability test for the 2-planet $N$-body linear ephemeris, has only
a formal sense, due to the large and unconstrained masses of the planets. We
selected a solution with semi-major axes $\sim 7$~au and relatively small
masses of the planets, $\sim 120$~$\mjup$, still unlikely in the real HU~Aqr
system. Similar to the previous experiment, stable models exhibit
$\Delta\lambda$ in anti-phase with the gathered statistics of the best-fitting
models. Stable systems with both semi-major axes $\sim 7$~au (close to the 1:1
MMR) are possible only when both planets are roughly aligned in their orbits
with {\em anti-aligned} apsidal lines, similar to the parabolic case (see the
left-hand panel in Fig.~\ref{fig:fig11}) and may persist even in extreme mass
ranges. Therefore, such apparently unlikely 1:1~MMR configurations {\em
should not be a'priori} considered as dynamically unstable. 

%
\subsection{Three-planet models with the linear ephemeris}
%
\label{sec:discussion}
The best-fitting 2-planet solutions NQ1 and NQ2 exhibit excessively large
period derivative $\beta \sim -10^{-12}~\mbox{day}\cdot L^{-2}$, which may be
interpreted as the LTT delay induced by a third companion with a very long
orbital period. This period may be estimated by the curvature of the parabolic
trend of the (O-C), compared to the observational window $\sim 20$~years.
Since only a small fraction of the orbit is apparent, the orbital period would
be multiplied by a factor of 2--3 or larger.

If to allow for a hypothesis of such a highly hierarchical 3-planet
configuration, we may focus first on the innermost 2-planet close-in systems.
In \citet{Gozdziewski2012}, we found stable $2$-planet configurations in
a~region of moderate eccentricities. However, similar to the HW~Vir case
\citep{Beuermann2012}, \corr{the outermost orbit of these models is not well
determined}. The present data of HU Aqr, extending the observational window by
only $\sim10\%$, seems to narrow possible solutions to a much more compact set
overlapping with the previously derived 3:2~MMR configurations. In the current
models, both semi-major axes are simultaneously shifted by $\sim 1$~au to a
secondary region of unstable models in the ($a_{\idm{b}},a_{\idm{c}}$)--plane
in \citet[][see their Fig.~A3, {\em left-bottom} panel]{Gozdziewski2012}.
Moreover, the best-fitting $N$-body configuration corresponds to the 1:1~MMR
with very similar orbital periods and masses, reaching the non-physical region
of two red dwarf companions. This solution is robust, and is preserved with
the up-to date midegress timing. The change of the orbital parameters is
caused by the fast decay of the (O-C) over past 2~years. We did not find any
{\em stable, coplanar 2-planet} strictly consistent with observations,
however, such systems might be stable if the planets were placed initially in
anti-phase with their actual, predicted ``observational'' positions. It may be
understood as a~strong, dynamical indication of plainly inadequate or
incomplete LTT model of the low-period component of the (O-C).

Therefore, we performed  the $N$-body search in terms of the 3-planet linear
ephemeris model, at first with planets in coplanar and direct orbits. As the result, we
obtain that basically any combination of masses up to the red dwarf mass
limit, semi-major axes in the range up to $20$~au, and eccentricities $\in
[0,1)$ are possible, still providing $\Chi \sim 1$. A~typical semi-major axis
of the outermost planet in these 3-planet models is larger than 16~au. We also
did not find any stable models with coplanar and direct orbits providing
$\Chi<3.3$, through setting different constraints for these models (moderate
eccentricities, circular orbit of the outermost component, hierarchical
configuration of the planets).

The assumption of coplanar systems with all direct orbits still does not
preclude stable {\em spatial} 2-- or 3--planet configurations, with high
mutual inclinations, up to the limit of coplanar counter-rotating companions.
In the most extreme case, such systems tend to be much more stable than
configurations with the direct orbits
\citep{Cuntz2010,Morais2012,Gozdziewski2013}. Strongly resonant systems may be
also possible, with three planets involved in deep, low-order and very narrow
MMRs, like in the HR~8799 planetary system exhibiting Laplace 4:2:1~MMRs
\citep{Gozdziewski2014} confined to small stable islands in the semi-major
axes--eccentricity space.

\begin{figure}
\centerline{   
   \hbox{\includegraphics[width=0.49\textwidth]{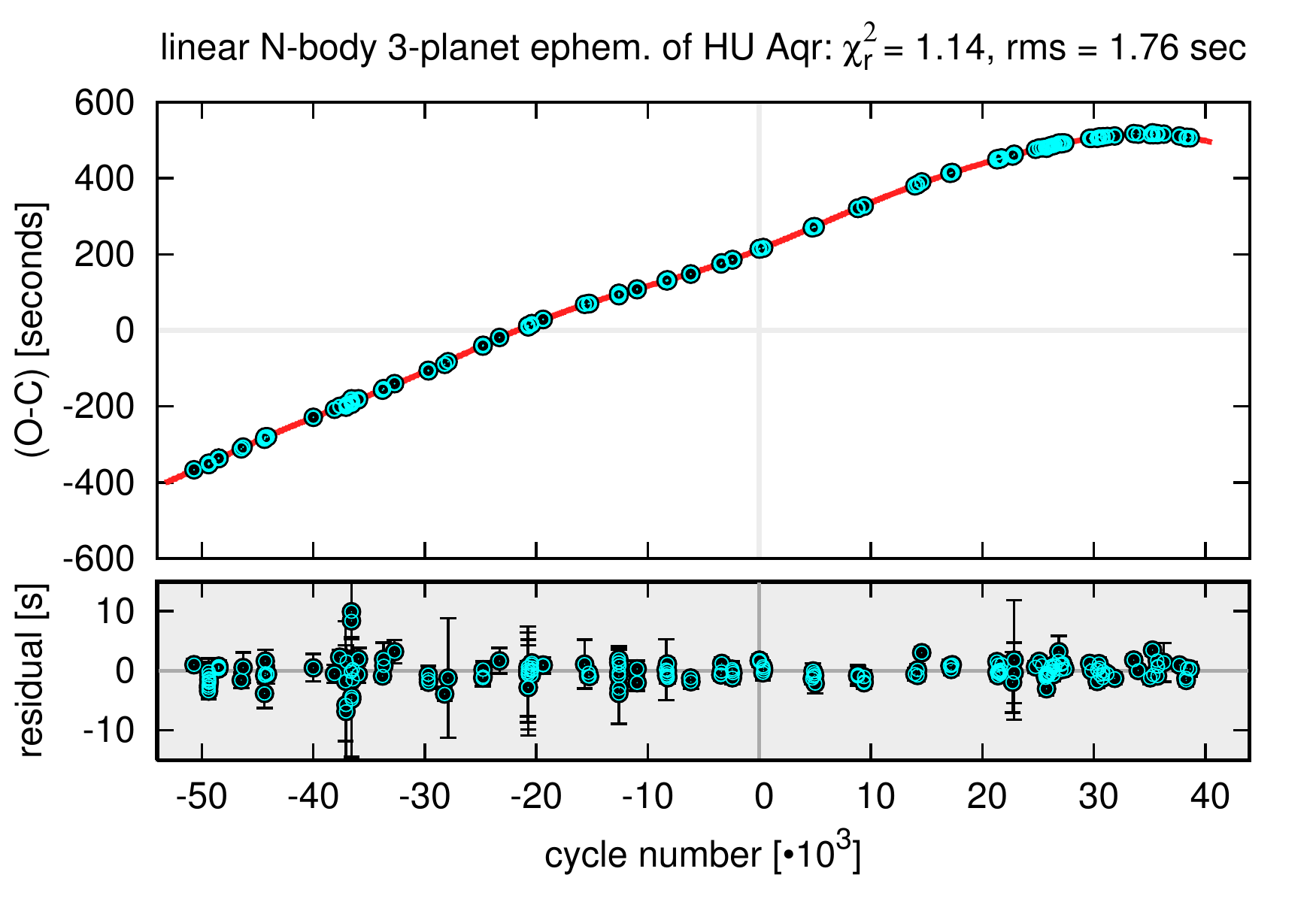}}   
}
\caption{
Synthetic curve of dynamically stable 3-planet model with
the linear ephemeris (NL3). The osculating,
astrocentric orbital elements at the epoch of
$T_0=$BJD~2,453,504.8882940, expressed by tuples
($m$ -- mass $[\mjup$], $a$ -- semi-major axis~[au], $e$ -- eccentricity, 
$i$ -- inclination~[deg],
$\Omega$ -- node~[deg], $\omega$ -- argument of pericenter~[deg], 
$M$ -- mean anomaly~[deg]) for each planet are:
(4.758~$\mjup$,  3.602~au,  0.0218, 90$^{\circ}$, 0$^{\circ}$,
173.01$^{\circ}$, 133.69$^{\circ}$),
(20.20~$\mjup$,  6.557~au,  0.1365, 90$^{\circ}$, 180$^{\circ}$,
10.77$^{\circ}$,  245.50$^{\circ}$), 
(80.00~$\mjup$, 12.887~au,  0.0158, 90$^{\circ}$,   0$^{\circ}$,
311.94$^{\circ}$,   83.47$^{\circ}$), for planets b, c, and~d, 
respectively. The initial epoch of the ephemeris $t_0$=BJD~2,453,504,888294,
$P_{\idm{bin}}$=0.0868202747~days. Total mass of the binary is 0.98~$M_{\sun}$.
This model provides $\Chi \simeq 1.14$ and an rms $\simeq 1.76$~s
with 205 data points and
17 free parameters (inclinations and nodes are fixed). The error
correction is $\sigma_{\idm{f}}$=0.9~seconds.
Formal uncertainties 
of the NL3 solution are difficult to estimate, since this 
model is displaced from the minimum of $\Chi
\sim 0.9$ and an rms $\sim 1.71$~second and is localised in dynamically
complex zone.
%
}
\label{fig:fig12}
\end{figure}

\begin{figure}
\centerline{   
   \hbox{\includegraphics[width=0.49\textwidth]{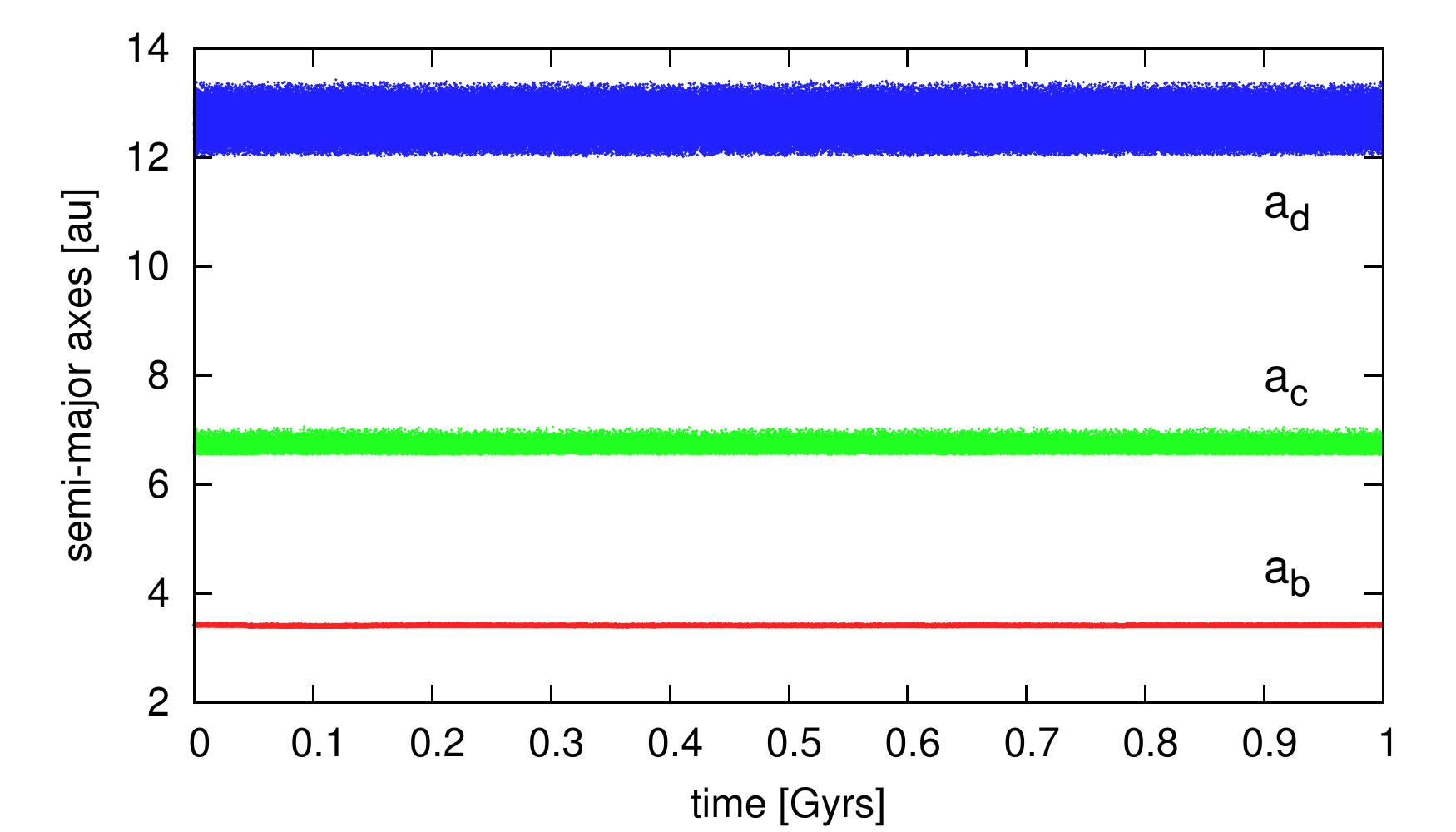}}   
}
\caption{
Evolution of semi-major axes in the long-term stable model NL3 with the middle
planet in a retrograde orbit. Osculating elements of this model are displayed
in caption to Fig.~\ref{fig:fig13}.
}
\label{fig:fig13}
\end{figure}

\begin{figure}
\centerline{   
   \hbox{\includegraphics[width=0.49\textwidth]{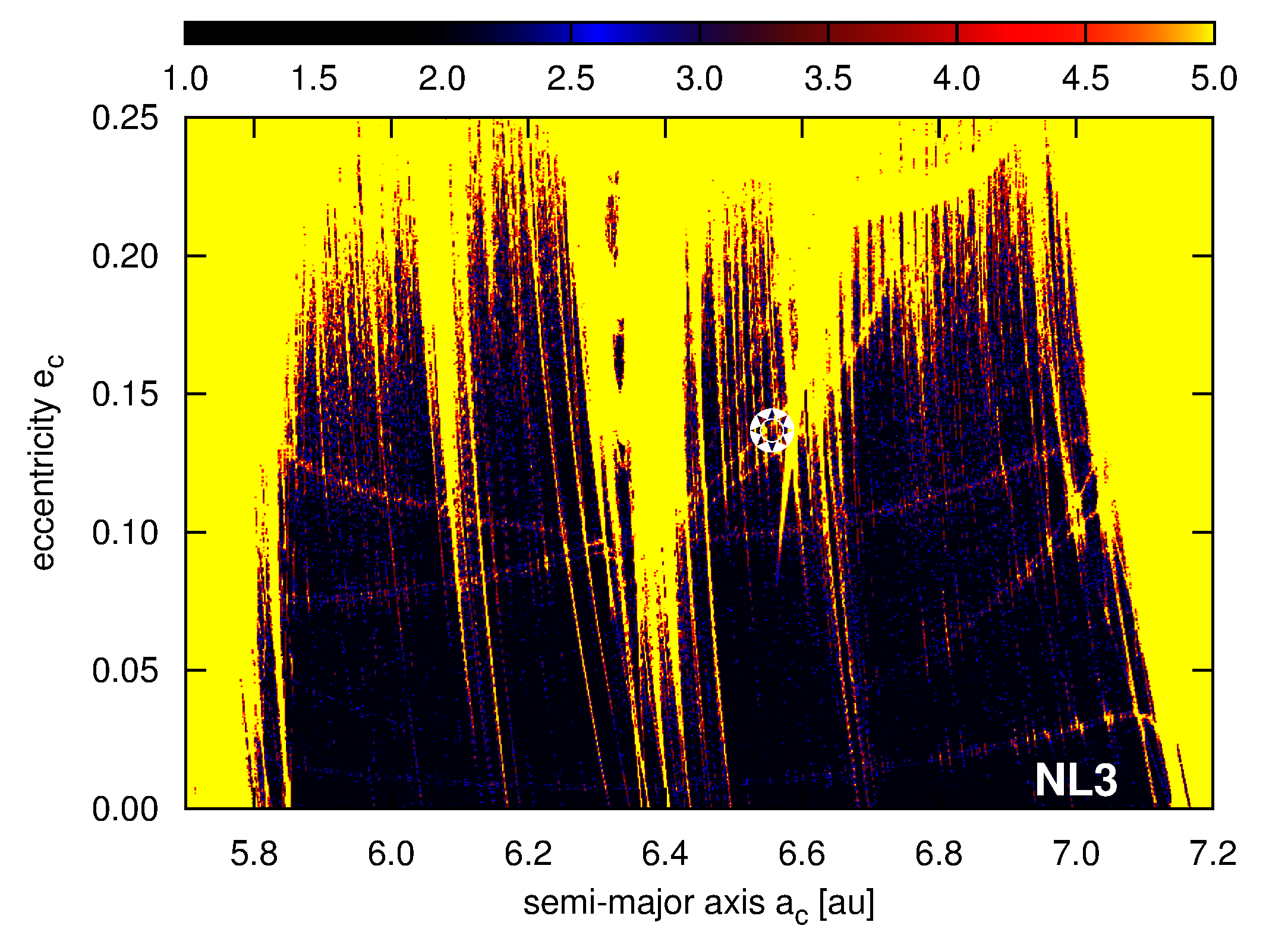}}   
}
\caption{
A dynamical map in terms of the MEGNO indicator for the long term stable model
NL3 in the (semi-major axis $a_{\idm{c}}$, eccentricity $e_{\idm{c}}$)--plane
of the middle planet. The nominal model is marked with the star symbol. The
dynamical stability is colour-coded: dark blue (black) with MEGNO $\sim 2$ is
for quasi-periodic configurations, yellow (grey) is for chaotic, strongly
unstable models. The resolution of the map is $1024\times768$ initial
conditions. Each configuration was integrated for 1.6~Myrs ($50,000$ outermost
periods). See the text for details.
}
\label{fig:fig14}
\end{figure}

As an illustration, we show here an example of a stable 3-planet model with
the linear ephemeris (NL3, see Fig.~\ref{fig:fig12}), which was found through
the hybrid algorithm, permitting that one of the planets is in {\em
retrograde} orbit with respect to two remaining planets in {\em prograde}
orbits. This solution, providing \corr{$\Chi\sim 1.14$ and very small} rms $1.76$~s
(similar to the best-fitting 2-planet solutions), belongs to a family of
stable, small-eccentricity orbits with semi-major axes of $\sim 3.6$~au, $\sim
6.6$~au and $\sim 13$~au, respectively. Moreover, the outermost planet's mass
is heavily unconstrained and may be as large as $80$~$\mjup$. 

In spite that the innermost and the middle planets' are also massive ($\sim
5$~$\mjup$ and $\sim 16$~$\mjup$, respectively), and the system is dynamically
packed, it remains stable for at least 1~Gyr. The evolution of bounded
semi-major axes for such an interval of time is shown in Fig.~\ref{fig:fig13}. To show
the dynamical neighbourhood of this particular model, we computed the MEGNO
dynamical map, varying the semi-major axis and eccentricity of the middle
planet, and keeping other parameter fixed at it's nominal values. The
dynamical map (Fig.~\ref{fig:fig14}) reveals that this model (and the whole
family of solutions of this class) is localised in a relatively extended
stable zone, spanned by a dense net of 2-body and 3-body MMRs. As a
``side--effect'' of this experiment, we detected the so called Arnold web,
seen as \corr{weak, vertical ``X''--like} structures in Fig.~\ref{fig:fig14}. The Arnold web
emerges due to overlapping unstable MMRs and their branches
\citep[e.g.][]{Gozdziewski2013} in strongly perturbed planetary systems.

This NL3 solution is peculiar, because the middle orbit is retrograde (the
orbital spin vector is anti-parallel to the orbital spins of two remaining
companions). We constrained the NL3 model by restricting the absolute
inclinations to one plane and fixing nodal lines, since this minimises the
number of free-parameters required to describe the spatial model. Releasing
these constraints, we also found stable solutions with mutual inclinations
reaching $180^{\circ}$, which result in reasonably small rms of $\sim
1.8$~seconds. Such apparently extreme solutions should not be necessarily
excluded as non-physical and non-realistic. The dynamical environment of the
putative planetary system has strongly changed during the common envelope
phase \citep{Zwart2013}. This might forced a dynamical instability of the
system and close encounters which resulted in highly inclined orbits.
%
%
\section{{Discussion and conclusions}}
%
%
The HU~Aqr binary belongs to the class of evolved binaries after the
post-common envelope phase or compact binaries, which presumably host single
or multiple planetary companions
\citep[e.g.]{Marsh2014,Lee2013,Almeida2013,Lee2012,Potter2011,Beuermann2011,
Beuermann2010,Qian2010,Yang2010}.
These papers claim (explicitly, or implicitly) that the observed (O-C)
variability can be explained by the R\o{}mer effect. Circumbinary planets
exist, indeed, since they have been recently detected and confirmed
independently through photometry of the {\sc Kepler} telescope
\citep[e.g.][]{Orosz2012,Welsh2012,Welsh2014}.

However, the LTT hypothesis suffers from ambiquities regarding optimisation
methods, dynamical models of putative planetary systems, as well as phenomena
intrinsic to the binaries.

Recent studies of multiple-planet systems detected by the (O-C) technique
revealed that their orbits are close to low-order MMRs and strongly unstable,
for time-scales as short as thousands of years
\cite[e.g.][]{Lee2014,Hinse2014d,Horner2013,Wittenmeyer2013,Lee2013,Horner2012a}.
The only well documented exception seems to be the NN~Ser system
\citep{Beuermann2010,Marsh2014} which presumably hosts a long--term stable
system of two Jovian planets involved in the 2:1~MMR or the 5:2~MMR. Also the
2-planet HW~Vir system \citep{Lee2009,Beuermann2012} may be resonant and
stable, though the semi-major axis of the outer planet, and its mass are yet
not well determined. HU Aqr is perhaps one of the most enigmatic members of
this class of putative planetary systems. Its analysis provides interesting
clues, spanning observational aspects, through the proper modelling of the
(O-C), the dynamical evolution and stability and possible scenarios of their
formation \citep[e.g.][]{Zwart2013}. 

In this work we tested the LTT effect, which may be present in the HU Aqr, on
strict dynamical grounds. In the the very recent paper, \citet{Bours2014}
conclude that the (O-C) derived 2-planet and 3-planet systems around this
binary are unstable due to large eccentricities. \corr{We found that such} coplanar
configurations are unstable due to semi-major axes in similar ranges and large
and unconstrained planetary masses reaching the non-physical range of red
dwarfs, and particular relative phasing on the planets in their orbits.
Actually, the best-fitting models exhibit small and moderate
eccentricities. They are degenerated due to strong correlations between their
physical and geometrical parameters.

The HU Aqr data provide a clear example that the Keplerian and $N$-body
formulations of the LTT effect for multiple planetary systems may lead to
qualitatively different views on its parameter space. Observational windows of
the binaries are narrow relative to the putative orbital periods, and the
inferred masses of hypothetical planets are large, up to the brown dwarf and
the red dwarf limits, like in the \corr{SZ~Her system \citep{Hinse2012a}}. In such settings, the
conversion of model parameters between both reference frames \corr{may introduce}
significant deviations between synthetic signals derived from the Keplerian,
and osculating Newtonian elements. This may qualitatively modify the
statistics of best-fitting configurations constrained by the available data.
Moreover, the LTT models of HU Aqr suffer from strong correlations between
different parameters, which makes the problem of reliable optimisation of
these models even more complex. Similar correlations are reported by
\cite{Marsh2014} for the parabolic ephemeris of NN Ser.

\begin{figure}
\centerline{   
   \hbox{\includegraphics[ width=0.48\textwidth]{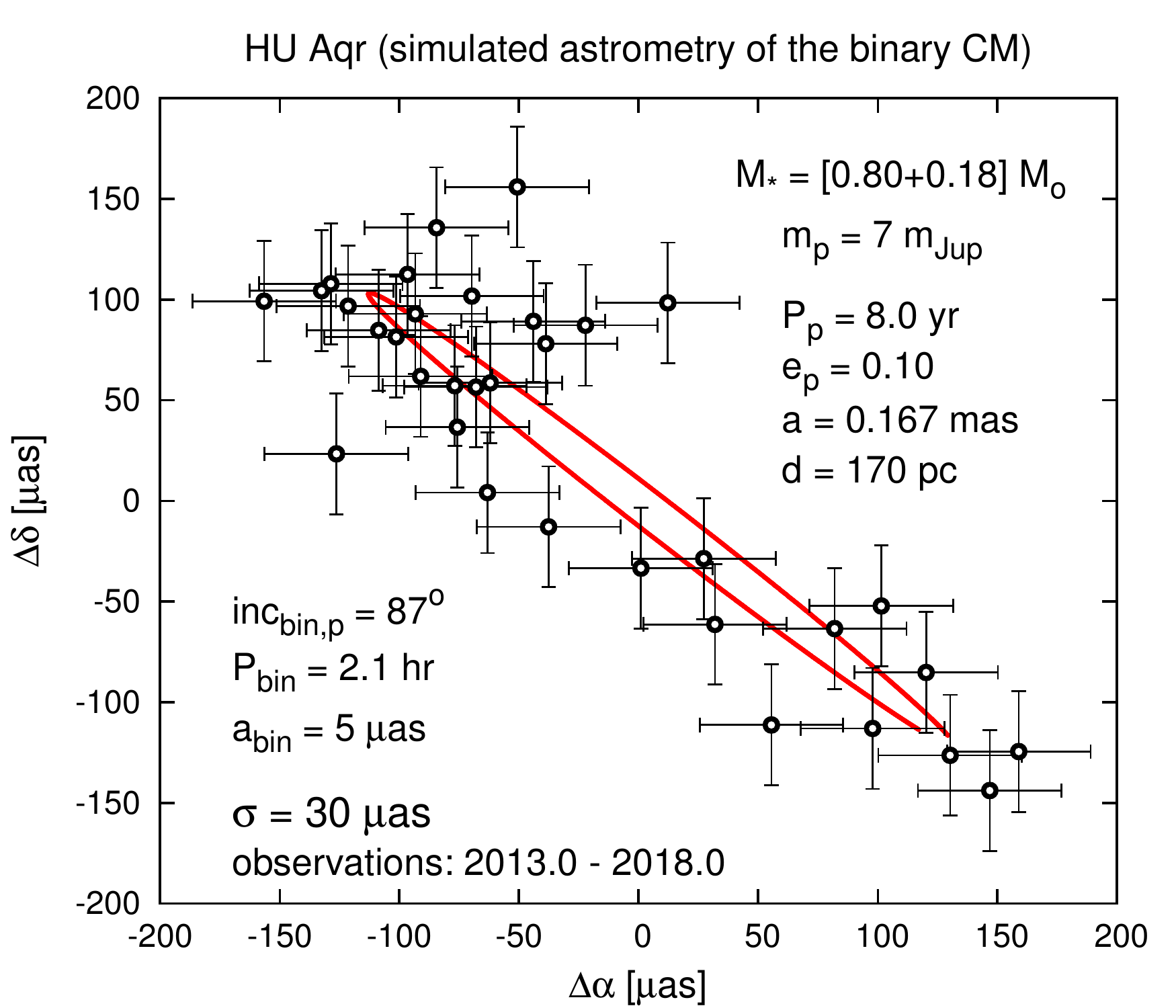}}   
}
\caption{
A simulation of the astrometric reflex motion of the mass centre (\corr{red
ellipse}) of the HU Aqr
(CM) due to the presence of a circumbinary 7 Jupiter mass planet at 8 years
orbit, \corr{coplanar with the binary orbital plane ($i=87^{\circ}$)}.  
The reflex semi-major axis $a$ of the CM is $\sim 0.17$~mas. Accurate
observations with the mean uncertainty at $30$--$50\,\mu$as level
(\corr{open circles with crossed error-bars}), and
spanning about of 5~years may be sufficient to detect the third-body object.
}
\label{fig:fig15}
\end{figure}

Our results indicate that the LTT hypothesis for the eclipse timing of HU~Aqr
is uncertain. Recalling the geometric source of the instability of coplanar
2-planet configurations, more elaborate dynamical models to describe the
observed (O-C) are required, like the 3-planet model with large relative
inclinations, as described above. 

Moreover, the results in \cite{Lanza1998,Lanza2006} still support the
quadrupole moment variations as the source of the (O-C), though the Applegate
effect is usually dismissed in the literature. To show this, we extrapolated
formulae in \citet[][their Eq. 7]{Wang2010} for the luminosity variation
$\Delta L/L$ of the secondary component M4V, as derived for the modified
Applegate mechanism in \citet[][and references therein]{Lanza1998,Lanza2006}.
We found that $\Delta L/L$ $\sim 8 \times 10^{-6}$ in the relative units,
adopting the radius $R=0.22 R_{\odot}$, mass $M = 0.18 M_{\odot}$, orbital
separation $a=0.0032$~au, and the effective temperature $T =3400$~K of the
secondary. Indeed, in accord with \cite{Lanza1998}, the luminosity variations
associated with the Applegate's mechanism should be effectively smoothed out,
given the much longer thermal timescale of the stellar convection zone and
they may be hardly observable. We also estimated the quadrupole period
variation $\Delta Q$ needed to drive the modulation of the orbital period
\citep{Lanza1999}
\[
 \frac{\Delta P_{\idm{bin}}}{P_{\idm{bin}}} = -9 \frac{\Delta Q}{M a^2} 
 \equiv \frac{4\pi K}{P_{\idm{mod}}},
\] 
where we adopted the semi-amplitudes $ K_{\idm{b,c}}$ and the orbital periods
$P_{\idm{b,c}}$ in \corr{Fit~JQ (Tab.~\ref{tab:tab3})} as estimates for the semi-amplitude $K$ of the
orbital period modulations, and for the modulation period $P_{\idm{mod}}$,
respectively. We obtain $\Delta P_{\idm{bin}}/P_{\idm{bin}} \sim 10^{-6}$,
$\Delta Q \sim 1 \times 10^{47}$~g\,cm$^2$ and $\sim 2 \times
10^{47}$~g\,cm$^2$, respectively, which are in the range characteristic for
other short-period CVs systems \citep{Lanza1999}. The order of magnitude of
these variations of $\sim 5 \times 10^{47}$~g\,cm$^2$ is typical for a
possible Applegate-like mechanism (A.~F.~Lanza, private communication).
Therefore, this mechanism leaves the LTT effect only as a reasonable
possibility among other explanations of the (O-C).

We certainly need an independent observational approach to shed more light on
this problem. Such a technique might be the astrometric monitoring of the
binary. HU Aqr is a relatively distant object at $\sim 200$~pc. However,
provided the accuracy of $\sim 30$~$\mu$s which is accessible by the ongoing
GAIA mission, a detection of its putative companions may be possible after
5~yr observations. A simulation of the astrometric signal of a single-planet
system with a $7$ Jovian mass planet in a $5$~au orbit (Fig.~\ref{fig:fig15})
reveals its amplitude of $\sim 0.3$~mas. If the putative outermost, massive
companion is very distant from the binary, as it might be suggested by the
$\beta$ term, then already well developed direct imaging technique might
confirm its presence directly. Combined with a proper dynamical model, such
a~detection might at least put some constraint on its mass and orbital
distance \citep{Gozdziewski2014}.

The results of our paper preclude the hypothesis of any coplanar 2-planet
system around HU Aqr. A 3-planet system with one planet in a retrograde orbit
or with high mutual inclinations might be possible, but its reliable detection
\corr{on the basis of the (O-C) data} is unlikely too, recalling the short time-span of the observations. It is also
very likely that the intrinsic binary's activity and the quadrupole
modulations of the secondary due to magnetic cycles may be fully responsible
for the observed (O-C) signal or for its significant fraction, similar to
stellar spots which ``pollute'' the Radial Velocities and transit data
\citep[e.g.][]{Montalto2014}. We postpone the analysis of such more complex
models of the (O-C) to future papers. The target should be systematically
monitored on a long-term time baseline to reveal the true nature of the
observed eclipse timing variability.
%
%
\section{Acknowledgements}
%
We thank the anonymous referee for thorough reviews, critical comments and
numerous, constructive suggestions that greatly improved this paper, and in
particular, for bringing the Lanza et al. modification of the Applegate
mechanism to our attention. We are very grateful to Antonino F. Lanza for
explanations regarding the Applegate effect and its observational constraints.
This work has been supported by Polish National Science Centre MAESTRO grant
DEC-2012/06/A/ST9/00276 (K.G.), SONATA grant DEC-2011/03/D/ST9/00656 (A.S.,
K.K., M.\.Z.), and SONATA BIS DEC-2011/01/D/ST9/00735 (M.G.). We thank the
Skinakas Observatory for their support and allocation of telescope time.
Skinakas Observatory is a collaborative project of the University of Crete,
the Foundation for Research and Technology – Hellas, and the
Max-Planck-Institute for Extraterrestrial Physics. 
This work is based on observations made with the 2-m telescope operated by the
National Astronomical Observatory in Rozhen (Bulgaria). We thank the staff of
NAO for their generous support and telescope time.
This work has made use of
data obtained at the Thai National Observatory on Doi Inthanon, operated by
NARIT. We also provide observations with the 1.55-m Carlos S\'anchez Telescope
operated on the island of Tenerife by the Instituto de Astrof\'isica de
Canarias in the Spanish Observatorio del Teide. We would like to thank Yucel
Kilic for his help with observations at the T\"UB\.ITAK National Observatory.
 K.G. thanks the Pozna\'n
Supercomputer and Network Centre (PCSS, Poland) for computational grant
No. 195 and technical support. Computations in this work were carried out on
the {\tt cane} and 
{\tt chimera} supercomputers of the PCSS.  \corr{
This research has made use of the SIMBAD database, operated at CDS,
Strasbourg, France, and of NASA's Astrophysics Data
System Bibliographic Services.}
\bibliographystyle{mn2e}
\bibliography{ms}
%
\appendix
\section{Mid-egress times of HU Aqr}
\label{sec:appendix}

Table~\ref{tab:tab5} in this Appendix collects mid-egress moments published in
the literature, as well as our new data, which were used for the analysis of
(O-C) in this paper.

The first column $L$ is the cycle number with respect to the epoch
BJD~2,449,102.9200026 (the BJD of the first observation in
\cite{Schwope1993}). The $L=0$ cycle here is shifted to epoch $T_0$=
BJD~2,453,504.8882940 with constant offset of 50702 cycles. The second column
is the moment of the mid-egress in MJD. The third column is the quoted
mid-egress error in the relevant source publication, derived on the basis of
observations with ``Instrument'' (fourth column). Instruments are code-named
in accord with a particular source publication: ``1'' is for
\citet{Schwope2001}, ``2'' is for \citet{Schwarz2009}, ``3'' is for
\citet{Qian2011}, ``4'' is for \citet{Gozdziewski2012}, ``5'' is for
\citet{Bours2014}, and ``6'' is for this work. The mid-egress measurements in
\citet{Qian2011} are included in Table~\ref{tab:tab5}, however, these data
were not used for constraining our models, since they differ by $\sim10$~s
from accurate OPTIMA and MONET/N points, spanning the same observational
window \citet{Gozdziewski2012}. There are 215 data mid-egress JD measurements
in total, spanning 89340 cycles since $L=0$,
but only $N_{\idm{obs}}=205$ points were used in this paper.

\bigskip
\ednote{Table~\ref{tab:tab5} will be available on-line. In this version
of the astro-ph preprint, most of the data points 
are commented out in ms.tex,  to save space, paper and green trees \ldots. 
Please extract the data from the source file, if the data are required.
}
%
\begin{table}
\caption{Mid-egress moments available in the literature.}
\begin{tabular}{rrrll}
\hline
cycle $L$ & MJD & Error [day] & Instrument & Ref. \\ 
\hline
       0   & 49102.4200026 & 0.0000029 & ROSAT & 1 \\ 
      1319 & 49216.9361120 & 0.0000115 & MCCP & 2 \\ 
      1320 & 49217.0229220 & 0.0000115 & MCCP & 2 \\ 
      1321 & 49217.1097490 & 0.0000115 & MCCP & 2 \\ 
      1322 & 49217.1966010 & 0.0000231 & ESO1m & 2 \\ 
      1333 & 49218.1516100 & 0.0000231 & ESO1m & 2 \\ 
      1334 & 49218.2384390 & 0.0000231 & ESO1m & 2 \\ 
      1367 & 49221.1035010 & 0.0000231 & ESO1m & 2 \\ 
      1368 & 49221.1903190 & 0.0000231 & ESO1m & 2 \\ 
      1369 & 49221.2771480 & 0.0000231 & ESO1m & 2 \\ 
      2212 & 49294.4667944 & 0.0000013 & ROSAT & 1 \\ 
      2213 & 49294.5536119 & 0.0000031 & ROSAT & 1 \\ 
      2216 & 49294.8140780 & 0.0000024 & ROSAT & 1 \\ 
      2222 & 49295.3349966 & 0.0000024 & ROSAT & 1 \\ 
      2225 & 49295.5954591 & 0.0000012 & ROSAT & 1 \\ 
      2226 & 49295.6822824 & 0.0000018 & ROSAT & 1 \\ 
      4241 & 49470.6254248 & 0.0000109 & ROSAT & 1 \\ 
      4409 & 49485.2112814 & 0.0000276 & ROSAT & 1 \\ 
      6328 & 49651.8196284 & 0.0000267 & ROSAT & 1 \\ 
      6341 & 49652.9483283 & 0.0000066 & ROSAT & 1 \\ 
      \ldots& \ldots        & \ldots    & \ldots      \\ 
     50702 & 53504.3882940 & 0.0000056 & ULTRA-VLT & 2 \\ 
      \ldots& \ldots        & \ldots    & \ldots      \\
     77031 & 55790.2824571 & 0.0000039 & MONET/N & 4 \\ 
     77066 & 55793.3211556 & 0.0000077 & MONET/N & 4 \\ 
     77067 & 55793.4079841 & 0.0000055 & MONET/N & 4 \\ 
     77078 & 55794.3630179 & 0.0000065 & MONET/N & 4 \\ 
     77247 & 55809.0356564 & 0.0000025 & LT+RISE & 5 \\
     77546 & 55834.9949490 & 0.0000179 & WFC & 4 \\ 
     77557 & 55835.9499905 & 0.0000295 & WFC & 4 \\ 
     77789 & 55856.0922852 & 0.0000038 & MONET/N & 4 \\ 
     77802 & 55857.2209399 & 0.0000090 & MONET/N & 4 \\ 
     77823 & 55859.0441786 & 0.0000066 & MONET/N & 4 \\ 
%
%
     77902 & 55865.9029864 & 0.0000022 & LT+RISE & 5 \\
     78100 & 55883.0934038 & 0.0000022 & MONET/N & 4 \\ 
     80324 & 56076.1818394 & 0.0000022 & LT+RISE & 5 \\
     80485 & 56090.1598976 & 0.0000019 & LT+RISE & 5 \\
     81001 & 56134.9591635 & 0.0000071 & OPT-SKO & 6 \\ 
     81013 & 56136.0010300 & 0.0000058 & OPT-SKO & 6 \\ 
     81162 & 56148.9372694 & 0.0000059 & OPT-SKO & 6 \\ 
     81186 & 56151.0209490 & 0.0000020 & OPT-SKO & 6 \\ 
     81231 & 56154.9278501 & 0.0000034 & OPT-SKO & 6 \\ 
     81486 & 56177.5670248 & 0.0000008 & WHT+UCAM & 5 \\
     81531 & 56180.9739470 & 0.0000022 & LT+RISE & 5 \\
     81532 & 56181.0607721 & 0.0000006 & WHT+UCAM & 5 \\
     81910 & 56213.8788462 & 0.0000002 & WHT+UCAM & 5 \\
     82566 & 56270.8329602 & 0.0000053 & LT+RISE & 5 \\
     84275 & 56419.2088830 & 0.0000011 & LT+RISE & 5 \\
     84678 & 56454.1974374 & 0.0000017 & LT+RISE & 5 \\
     85746 & 56546.9214776 & 0.0000073 & PIVA-NAO & 6 \\ 
     85965 & 56565.9351702 & 0.0000010 & LT+RISE & 5 \\
     86032 & 56571.7520793 & 0.0000034 & PIVA-NAO & 6 \\ 
     86391 & 56602.9205819 & 0.0000016 & LT+RISE & 5 \\
     86412 & 56604.7437774 & 0.0000027 & PIVA-NAO & 6 \\ 
     86433 & 56606.5670216 & 0.0000045 & TNT+USPEC & 5 \\
     86467 & 56609.5189097 & 0.0000021 & TNT+USPEC & 5 \\
     86976 & 56653.7104280 & 0.0000361 & PW24-COG & 6 \\ 
     88383 & 56775.8665044 & 0.0000017 & ULTRA-TNT & 6 \\ 
     88973 & 56827.0904134 & 0.0000016 & INT+WFC & 5 \\
     88985 & 56828.5322517 & 0.0000037 & INT+WFC & 5 \\
     89066 & 56835.1647131 & 0.0000024 & LT+RISE & 5 \\
     89339 & 56858.8666325 & 0.0000064 & PIVA-NAO & 6 \\ 
     89340 & 56858.9534517 & 0.0000103 & PIVA-NAO & 6 \\ 
   \hline
\end{tabular}
\label{tab:tab5} 
\end{table}
\label{page:lastpage}
\end{document}